\documentclass[12pt]{article}

\usepackage{scicite}

\usepackage{times}
\usepackage{amssymb}
\usepackage{graphicx} 
\usepackage{todonotes} 
\usepackage{gensymb}
\usepackage{amsmath}
\usepackage[labelfont=bf]{caption}
\usepackage{mathtools}
\usepackage{xcolor}
\usepackage{physics}
\usepackage{bm}

\newcommand{\supplementarysection}{%
  \setcounter{figure}{0}
  \let\oldthefigure\thefigure
  \renewcommand{\thefigure}{S\oldthefigure}
  \setcounter{section}{0}
  \let\oldthesection\thesection
  \renewcommand{\thesection}{\oldthesection}
  \setcounter{equation}{0}
  \let\oldtheequation\theequation
  \renewcommand{\theequation}{S\oldtheequation}
  \setcounter{table}{0}
  \let\oldthetable\thetable
  \renewcommand{\thetable}{S\oldthetable}
}

\newcommand{\TNeel}{T_{\text{N}}}
\newcommand{\Bflake}{B_{\text{flake}}}

\newenvironment{sciabstract}{%
\begin{quote} \bf}
{\end{quote}}

\title{Quantum Noise Spectroscopy of Criticality in an Atomically Thin Magnet}

\author
{Mark E. Ziffer$^{1,2,3,\P,\dagger,*}$, Francisco Machado$^{4,5,6,\dagger,*}$, Benedikt Ursprung$^2$, \\Artur Lozovoi$^{7,8}$, Aya Batoul Tazi$^1$, Zhiyang Yuan$^7$, Michael E. Ziebel$^3$,\\ Tom Delord$^8$, Nanyu Zeng$^9$, Evan Telford$^3$, Daniel G. Chica$^3$,\\
Dane W. deQuilettes$^{10,11}$, Xiaoyang Zhu$^3$, James C. Hone$^2$, Kenneth L. Shepard$^9$, 
\\Xavier Roy$^3$, Nathalie P. de Leon$^7$, Emily J. Davis$^{12}$, \\ Carlos A. Meriles$^{8,*}$, Jonathan S. Owen$^{3,*}$, P. James Schuck$^{2,*}$, Shubhayu Chatterjee$^{13,*}$,\\ 
Abhay N. Pasupathy$^{1, 14,*}$\\
\normalsize{$^{1}$Department of Physics, Columbia University, New York, NY 10027 USA}\\
\normalsize{$^2$ Department of Mechanical Engineering, Columbia University, New York, NY 10027 USA}\\
\normalsize{$^3$ Department of Chemistry, Columbia University, New York, NY 10027 USA}\\
\normalsize{$^4$ ITAMP, Harvard-Smithsonian Center for Astrophysics, Cambridge, MA 02138, USA}\\
\normalsize{$^5$ Department of Physics, Harvard University, Cambridge, MA 02138, USA}\\
\normalsize{$^6$ QuTech and Kavli Institute of Nanoscience, Delft University of Technology,}\\
\normalsize{Delft, 2600 GA, The Netherlands}\\
\normalsize{$^7$ Department of Electrical and Computer Engineering, Princeton University,}\\ \normalsize{Princeton, NJ 08540, USA}\\
\normalsize{$^8$ Department of Physics, CUNY, The City College of New York, New York, NY, 10031, USA}\\
\normalsize{$^9$ Department of Electrical Engineering, Columbia University, New York, NY 10027 USA}\\
\normalsize{$^{10}$Lincoln Laboratory, MIT, Lexington, MA, 02421, USA}\\
\normalsize{$^{11}$Center for Quantum Engineering, MIT, Cambridge, MA, 02139, USA}\\
\normalsize{$^{12}$Department of Physics, New York University, New York, NY 10003, USA}\\ 
\normalsize{$^{13}$Department of Physics, Carnegie Mellon University, Pittsburgh, PA 15213, USA}\\
\normalsize{$^{14}$Condensed Matter Physics and Materials Science Department,}\\ 
\normalsize{Brookhaven National Laboratory, Upton, NY, 11973 USA}\\
\\
\normalsize{$\dagger$ These authors contributed equally}
\\
\normalsize{$^\ast$Corresponding author. Email:   apn2108@columbia.edu, p.j.schuck@columbia.edu,}\\ \normalsize{jso2115@columbia.edu, mark@qunett.com, cmeriles@ccny.cuny.edu,}\\  \normalsize{shubhayuchatterjee@cmu.edu}, 
\normalsize{f.machado@tudelft.nl}\\
\\
\normalsize{$\P$ Present address: Quantum Network Technologies, Inc., Boston, MA 02215}
}

\date{}

\begin{document}

\maketitle 

\newpage

\begin{sciabstract}
The universality class of a phase transition dictates the behavior of low-frequency and long-wavelength critical fluctuations in a material.
Studying these dynamics in thin magnetic insulators, however, remains challenging; an experimental probe must be sensitive to low-frequency, long-wavelength fluctuations while navigating the sample's small volume, inherent inhomogeneities, and fabrication limitations.
Using nitrogen-vacancy (NV) centers in diamond, we demonstrate $T_2$ (spin-decoherence) noise magnetometry as a direct and quantitative probe of critical dynamics in a tri-layer sample of CrSBr, a two-dimensional van der Waals magnet. 
We observe a sharp increase in the decoherence rate of the NV spin ensemble at the sample's critical temperature, highlighting our probe's sensitivity to critical fluctuations. We demonstrate that the behavior of the NV spin decoherence rate below the critical temperature is inconsistent with fluctuations due to XY criticality in 2D CrSBr.
Next, by simulating the NV spin decoherence dynamics using noise models for  an Ising-type transition and fitting to our data, we quantitatively measure the dynamical and the correlation length critical exponents ($z$ and $\nu$) and demonstrate that the observed fluctuations strongly deviate from those predicted by the short-range Ising universality class. Together, these observations indicate that the transition is Ising-like but modified by long-range dipolar interactions, emphasizing the importance of easy-axis anisotropy and long-range interactions for understanding magnetism in 2D CrSBr. 
Our work establishes $T_2$ noise magnetometry as a quantitative probe of critical dynamics in 2D materials, opening a new avenue for studying strongly correlated phenomena at the atomically thin limit.

\end{sciabstract}

\section*{}

Near criticality, the macroscopic properties of a material are determined by the dimensionality and symmetry of the underlying microscopic degrees of freedom ~\cite{cardyScaling1996, hohenbergTheory1977}. 
This observation lies at the heart of universality: distinct materials and physical systems can exhibit the same scaling behavior near their critical points, provided that the underlying interactions exhibit the characteristics of a particular universality class. 
By carefully characterizing the low-frequency and long-wavelength fluctuations of a material, one can build a catalog of scaling functions and critical exponents and their corresponding universality classes.
On one hand, such analysis offers a better understanding of the relevant features of the microscopic interactions that lead to order in different materials, such as principal anisotropies, effective dimensionalities, and interaction range. 
On the other hand, such characterization immediately informs us of the behavior of all physical systems in the same universality class~\cite{cardyScaling1996}.

In magnetic materials, the experimental characterization of fluctuations has played a fundamental role in understanding magnetic ordering in quantum materials, and in establishing the universality class of critical phenomena~\cite{freyCritical1994, vojtaQuantum2003}. 
The recent discovery of stable magnetic phases in atomically thin van der Waals (vdW) crystals has greatly expanded the landscape of intriguing magnetism, bringing forth the opportunity to study bona fide two-dimensional phenomena~\cite{ahnProgress2024}. 
However, probing atomically thin vdW magnets poses a challenge for conventional magnetic probes: small sample volumes limit the use of conventional techniques such as inelastic neutron scattering, $\mu$SR, NMR, and EPR for studying magnetic dynamics \cite{freyCritical1994, machadoQuantum2023}.
These constraints have motivated the use of alternative approaches and techniques, including quasi-elastic light scattering and magneto-optical techniques \cite{lujanMagnons2022, jinImaging2020,zhangSpin2021} or sample preparation approaches involving aggregation of 2D flakes to increase the  sample volume \cite{mahatoDimensiondependent2024}.
However, these approaches either require complex sample preparation schemes or are limited to probing
correlations at length scales on the order of the diffraction limit of an optical probe.

Recently, noise spectroscopy utilizing solid-state spin defects has been proposed as a quantitative probe of critical fluctuations in mesoscopic 2D magnets\cite{machadoQuantum2023}.
The ability to isolate, initialize, and control nano-scale spin defects near a sample of interest ensures that the sensitivity of the probe is set by its distance relative to the sample, thus overcoming the need for large sample volumes.
Furthermore, the sample-probe distance serves as an important experimental tuning parameter: a spin qubit probe placed a distance $\ell$ away from the two-dimensional material is selectively sensitive to phenomena at wavenumber $\sim 1/\ell$\cite{machadoQuantum2023}, enabling the direct probing of long-wavelength correlations on length scales below the optical imaging resolution.
Finally, spin qubit noise spectroscopy offers a wide dynamical range to probe fluctuations  ($\sim\mathrm{kHz} - \mathrm{GHz}$) depending on whether depolarization ($T_1$) or decoherence ($T_2$) noise sensing protocols are employed ~\cite{liuQuantum2025, xueMagnon2026}.

$T_1$ noise sensing with nitrogen-vacancy (NV) spin defects in diamond has previously been leveraged to study fluctuations in condensed matter systems \cite{duControl2017,hsiehImaging2019, andersenElectronphonon2019, wangNoninvasive2022} and even probe magnetic phases in 2D materials \cite{duControl2017, wangNoninvasive2022,huangRevealing2023, melendezQuantum2025}.
However, frequency resolution in this technique is achieved by applying and controlling an external bias field.
To avoid introducing large static fields that perturb the magnetic response of a target material, $T_1$ noise sensing experiments on magnetic materials are typically limited to probing fluctuation dynamics near the spin's zero-field splitting frequency (in the $\sim$ few $\mathrm{GHz}$ range for the NV center).
In contrast, the target frequency in $T_2$ noise spectroscopy is fully determined by the applied pulse sequence~\cite{machadoQuantum2023}, which enables access to much lower frequency ranges ($\sim\mathrm{kHz}$-$\mathrm{MHz}$) at small, fixed external fields.
This has enabled the use of NV $T_2$ noise spectroscopy to probe low-frequency phenomena as diverse as nuclear and electron spin-bath dynamics in diamond, low-energy dynamics of 2D magnets, and NMR of single-molecule targets\cite{laraouiHighresolution2013,staudacherNuclear2013,myersProbing2014,rosskopfInvestigation2014,schafer-nolteTracking2014,tetienneScanning2016,dwyerProbing2022,davisProbing2023, xueMagnon2026, liCritical2025}.

In this work, we utilize $T_2$ noise spectroscopy with NV spin ensembles to comprehensively study low-frequency, long-wavelength critical magnetic fluctuations in a tri-layer sample of the 2D vdW magnet CrSBr \cite{ziebelCrSBr2024}~(Fig.~\ref{fig:fig1}).
Critical fluctuations at the magnetic ordering transition in CrSBr are revealed by a sharp increase in the NV decoherence rate around the magnetic ordering temperature~(Fig.~\ref{fig:fig2}).
Crucially, the $T_2$ decoherence profile encodes important details about the two-point correlation function that governs the critical fluctuations, from which we conclude that the critical behavior is not consistent with either the XY or short-range Ising models.
While XY criticality predicts a broad, highly asymmetric decoherence rate near the transition temperature, we observe a sharp feature~(Fig.~\ref{fig:fig3}).
Moreover, we quantitatively measure  the dynamical and correlation length critical exponents ($z$ and $\nu$) and find values that are not consistent with those predicted for standard short-range Ising universality (Fig.~\ref{fig:fig3}); this stark difference emphasizes the role of long-range dipolar magnetic interactions in 2D CrSBr~\cite{sunDipolar2024}. 
These results both clarify outstanding questions about the nature of the magnetic transition in CrSBr at the two-dimensional limit and establish NV $T_2$ noise spectroscopy as a precise technique to study criticality in 2D materials.

\section*{Probing Magnetic Ordering in 2D CrSBr}

\begin{figure*}
\includegraphics[width=1\textwidth]{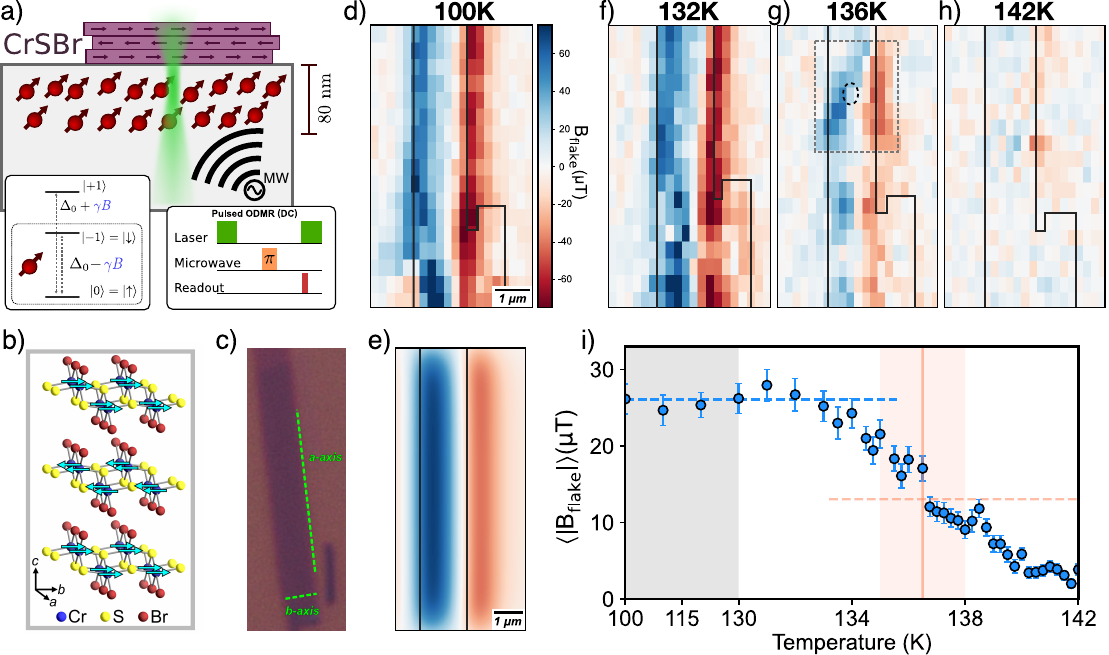}

\caption{\textbf{Magnetic imaging of 2D CrSBr.}~
{\bf (a)} Schematic of the experimental configuration: a tri-layer CrSBr flake  transferred onto a diamond substrate with an NV ensemble (red spins) implanted at a shallow depth below the surface \cite{SM}. 
A focused green laser beam and a microwave (MW) source depict optical addressing and microwave delivery, respectively \cite{SM}. 
\emph{(Inset, bottom left)} Level structure of the $S=1$ spin ground state of the NV center, where $\Delta_0$ and $\gamma B$ refer to the zero field splitting and Zeeman energy respectively. 
\emph{(Inset, bottom right)} Pulse sequence for the pulsed ODMR DC magnetometry protocol \cite{SM}. 
{\bf (b)} Atomic and magnetic structure of a CrSBr tri-layer, with turquoise arrows representing the magnetic moment on Cr atoms. 
{\bf (c)} Brightfield optical image of the tri-layer CrSBr flake on a SiO$_2$($285 \:\mathrm{nm}$)/Si substrate. 
{\bf (d)} Stray field image of the tri-layer CrSBr flake at $100\:\mathrm{K}$ constructed from a confocal scan of the ODMR measurement. An outline of the flake based on co-registration with a PL image is shown by the black lines \cite{SM}. 
{\bf (e)} Simulation of the stray field projection from a single uncompensated FM layer of CrSBr onto the depth distribution of our NV ensemble \cite{SM}. The black outline represents the position of the simulated CrSBr flake.
{\bf (f-h)} Stray field images of tri-layer CrSBr for different temperatures across the magnetic phase transition ($132$, $136$, and $142$~K). The small dashed circle in (g) indicates the region on the flake where spin-echo measurements are performed. The dashed box indicates the region of the image taken to calculate the average absolute field $\langle|\Bflake|\rangle$ \cite{SM}. 
{\bf (i)} $\langle|\Bflake|\rangle$ as a function of temperature spanning a broad low temperature range (grey) and a narrow range across the phase transition. 
The transition highlighted corresponds to the temperature where $\langle|\Bflake|\rangle$ is half of its low temperature value (blue dashed line).}  
\label{fig:fig1}
\end{figure*}

CrSBr is a vdW semiconductor with A-type layered antiferromagnetic (AFM) ordering below the N\'{e}el temperature $\TNeel\sim132 \:\mathrm{K}$ for bulk samples~\cite{lopez-pazDynamic2022, telfordCoupling2022, leeMagnetic2021, scheieSpin2022}.
Below $\TNeel$, the intralayer magnetic order is ferromagnetic (FM) with in-plane easy-axis ordering along the crystallographic $b$-axis.
Along the $c$-axis, adjacent layers magnetize along opposite directions leading to interlayer AFM order (Fig.~\ref{fig:fig1}b). 
In the thin film regime, a sample composed of an odd number of layers exhibits an overall net magnetization arising from one uncompensated FM layer~\cite{rizzoVisualizing2022, tschudinImaging2024}.
We focus our study on a tri-layer CrSBr flake whose edges align with the crystallographic $a$ and $b$ axes (Fig.~\ref{fig:fig1}c).

We probe the CrSBr flake by transferring it onto an isotopically pure $^{12}\mathrm{C}$ diamond substrate with an ensemble of  NV spin defects implanted near the surface ~(Fig.~\ref{fig:fig1}a)~\cite{SM}. 
Each NV defect hosts an $S=1$ spin ground state manifold that can be optically initialized and read out, and coherently manipulated using microwaves~\cite{dohertyNitrogenvacancy2013}.
The $\ket{m_s=0}$ and  $\ket{m_s=\pm 1}$ levels are split by a $2.87$ GHz zero field splitting, and the degeneracy of $\ket{m_s=\pm 1}$ is broken by applying a small external bias field ($\sim3.5$~mT) along the NV spin quantization axis \cite{SM}.
This isolates a two-level spin-qubit defined by the $|m_s=0\rangle = \ket{\uparrow}$ and $\ |m_s=-1\rangle = \ket{\downarrow}$ states (Fig.~\ref{fig:fig1}a).
While our imaging resolution is determined by the diffraction limit of our confocal laser scanning setup ($\sim 500$ nm), each individual NV spin is sensitive to fluctuations whose wavelength approximately matches the NV-sample distance ($\ell \in [10,80]$ nm)~\cite{SM}. Note that this length scale is smaller than the domain size reported in previous literature ~\cite{tschudinImaging2024}.
The observed signal corresponds to an integration over the depth distribution of NV centers within a confocal spot~(Fig.~\ref{fig:fig1}a)~\cite{SM}.

We first study the static magnetic properties of the system using pulsed optically detected magnetic resonance (ODMR) (Fig.~\ref{fig:fig1}a).
By converting the measured change in spin resonance frequency to the static magnetic field projection along the NV spin quantization axis, we can directly image the stray magnetic field  generated by the tri-layer CrSBr flake, $\Bflake$,  to  study the formation of magnetic ordering~\cite{SM, bstray}.
When cooling the sample to $100$~K (Fig.~\ref{fig:fig1}d) we observe a dipolar pattern for $\Bflake$ that matches the field generated by the in-plane magnetization of a single layer in the CrSBr flake (Fig.~\ref{fig:fig1}e), providing evidence for the expected intra-layer ferromagnetic and inter-layer anti-ferromagnetic ordering~\cite{lopez-pazDynamic2022,rizzoVisualizing2022,tschudinImaging2024, ghiasiNitrogenvacancy2023, SM}.

Next, we study the melting of the magnetic ordering by imaging $\Bflake$ as a function of increasing temperature.
Starting from a uniformly magnetized flake below $132$~K (Fig.~\ref{fig:fig1}f), the sample begins to develop $\mathrm{\mu m}$-scale magnetic field patterns at $\sim136$~K, capturing an inhomogeneous loss of magnetic order across the flake (Fig.~\ref{fig:fig1}g).
These domains are present until the temperature is raised to $\sim 142$~K (Fig.~\ref{fig:fig1}h), at which point $\Bflake$ becomes negligible, signaling the loss of magnetic ordering in the CrSBr flake.
The observed magnetic domains are consistent with previous magnetic imaging studies on 2D CrSBr, where domain formation has been attributed to inhomogeneous strain that results in a spatially varying interlayer exchange coupling strength~\cite{rizzoVisualizing2022,tschudinImaging2024}.

To quantify the loss of magnetic order and better locate the magnetic phase transition, we define a region of pixels in one large-scale domain (square in Fig.~\ref{fig:fig1}g) and measure the average absolute value of $\Bflake$ as a function of temperature~(Fig.~\ref{fig:fig1}i).
From the loss of magnetization we identify a phase transition temperature in the range of $\sim135-138$~K corresponding to the reduction of $\langle |\Bflake |\rangle$ by half of its maximal value.
This broad feature is consistent with previous studies on the magnetic transition in few-layer CrSBr~\cite{ziebelCrSBr2024, tschudinImaging2024}.

\section*{Measuring Critical Fluctuations with $T_2$ Noise Spectroscopy}

\begin{figure*}
\centering
\includegraphics[width=0.7\linewidth]{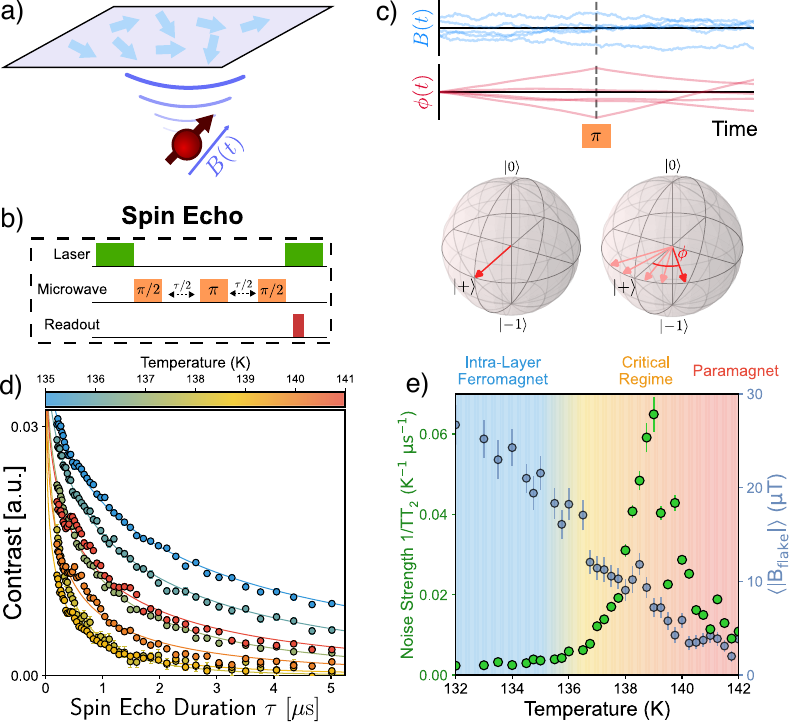}
\caption{\textbf{Critical fluctuations in CrSBr observed with NV $T_2$ noise measurements.}
{\bf (a)} Fluctuating magnetic moments in CrSBr induce a time-varying magnetic field $B(t)$ along the quantization axis of an NV probe spin. 
{\bf (b)} Spin-echo pulse sequence used to measure the NV's coherence.
{\bf (c)} Over each experimental shot $B(t)$ is random; consequently, the phase $\phi(t)$ accumulated by the NV is also random.
The decoherence decay encodes the growth of the variance of $\phi(\tau)$ averaged over experimental shots, $e^{-\langle\phi^2(\tau)\rangle/2}$. 
{\bf (d)} Spin-echo coherence decays measured at the position indicated in Fig.~\ref{fig:fig1}g for several representative temperatures across the CrSBr magnetic phase transition, fitted to a stretched exponential to extract the decoherence rate $1/T_2$ \cite{t2,SM}. 
Using an additional analysis of the $\pi-$pulse contrast \cite{SM}, we observe that the initial ($\tau\to 0$) spin-echo contrast approaches a common value for all measurements taken over this temperature range.
Based on our Rabi frequency, the shortest experimental $\tau$ is $200$~ns.
{\bf (e)} Comparison between $\langle|\Bflake|\rangle$ and $1/TT_2$ as a function of temperature across the CrSBr phase transition. The critical temperature $T_C=139~\mathrm{K}$ is identified from the peak position of $1/TT_2$.}
\label{fig:fig2}

\end{figure*}

Having characterized the magnetic ordering in 2D CrSBr using static magnetic imaging, we next investigate the dynamical features of the magnetic phase transition by studying its low-frequency magnetic fluctuations.
The key insight is that spin fluctuations in the CrSBr flake lead to fluctuations of the magnetic field ($B(t)$) at the NV location; by characterizing these magnetic field fluctuations using $T_2$ noise sensing with our NV ensemble we can directly probe the dynamical properties of CrSBr across the magnetic phase transition~(Fig.~\ref{fig:fig2}a). We implement this approach by  measuring the coherence decay of the NV ensemble at a position directly beneath the CrSBr flake (Fig.~\ref{fig:fig1}g) as a function of temperature using a spin-echo pulse sequence.

The spin-echo measurement (Fig.~\ref{fig:fig2}b) proceeds as follows:
after optical initialization into $\ket{\uparrow}$, a $\pi/2$ rotation prepares the NV spins in the superposition $\ket{+} = (\ket{\uparrow} + \ket{\downarrow})/\sqrt{2}$.
The spin evolves for time $\tau/2$ under the external fluctuating field, a $\pi$-pulse is applied, and the system evolves for another time $\tau/2$ into $(\ket{\uparrow} + e^{-i\phi}\ket{\downarrow})/\sqrt{2}$.
A final $\pm \pi/2$-pulse is applied to map the accumulated phase $\phi$ of the resulting state into a population (\{$\ket{\uparrow}$,$\ket{\downarrow}$\}) that can be optically read out to determine a population difference proportional to $\text{Re}[e^{-i\phi}]$.
Integrated over experimental shots and the ensemble of NV centers within the area of the confocal optical probe, the observed signal $C(\tau)$ captures the average of the coherence ($\langle \text{Re}[e^{-i\phi}]\rangle$). 

Crucially, this measured coherence $C(\tau)$ serves as a frequency-selective probe of fluctuations in the external field. Under the spin-echo protocol, each spin qubit precesses by an angle $\phi(\tau) = -\gamma\int_0^{\tau/2} dt~B(t) + \int_{\tau/2}^\tau dt~B(t)$ which measures the \emph{difference} in the time-averaged local magnetic field before and after the $\pi$-pulse. 
The interaction between the spin qubit and the magnetic environment is modulated under this protocol such that  interactions between the qubit and (near)-static or fast oscillating  magnetic fields are decoupled,  maximizing the qubit's sensitivity to fluctuations of $B(t)$ at frequencies $\sim 1/\tau$. 
Over different experimental shots, each NV experiences a different random $B(t)$ and thus precesses over a random phase $\phi(\tau)$ (Fig.~\ref{fig:fig2}c). 
The growth of fluctuations of $\phi(\tau)$ then determines the experimental-shot-averaged, single-NV decoherence dynamics for each NV in the ensemble: $\langle \text{Re}[e^{-i\phi}]\rangle \approx e^{-\langle\phi^2(\tau)\rangle/2}$~\cite{desousaElectron2009,machadoQuantum2023}.

While the natural energy scale of magnetism in CrSBr is much larger than $1/\tau$, the characteristic frequency of its magnetic fluctuations is greatly reduced as the phase transition is approached; this phenomenon of critical slowing down is expected at continuous phase transitions~\cite{hohenbergTheory1977}. As a result, we expect to observe an enhancement of the  measured decoherence rate as the NV ensemble interacts with low frequency critical fluctuations in CrSBr, making the temperature dependence of the decoherence rate a sensitive probe of the location of the phase transition \cite{machadoQuantum2023}. This prediction is borne out by the data: the spin-echo coherence dynamics exhibit a fast decay when the sample temperature approaches the critical temperature (Fig.~\ref{fig:fig2}d). Focusing on the temperature-weighted decoherence rate $1/TT_2$ \cite{t2, invtt2}, we observe a sharp enhancement near 139~K (Fig.~\ref{fig:fig2}e), indicating the presence of a continuous phase transition that is consistent with the trend in $\langle|\Bflake|\rangle$ (Fig.~\ref{fig:fig1}i).
Moreover, the sharp peak in the decoherence rate as a function of temperature allows us to assign a precise and local measurement of the critical temperature $T_C=139$~K.

\section*{Analysis of the Critical Dynamics}
The decoherence dynamics measured by our NV ensemble contain far more information than the location of the transition alone; a closer analysis of their temperature dependence reveals information about the universality class governing the critical behavior of CrSBr \cite{machadoQuantum2023}.
The temperature dependence of the decoherence rate $1/TT_2$ by itself encodes important features of the evolution of magnetic fluctuations with temperature that can help us constrain the nature of the phase transition. 
Two hypotheses have been put forward to explain the magnetic phase transition in CrSBr:
the system undergoes an Ising transition into a long-range ordered easy-axis ferromagnet,
or the system undergoes a BKT transition into a quasi-long-range ordered 2D XY magnet ~\cite{ziebelCrSBr2024}.
These two hypotheses predict different power-law exponents for the onset of the sample's static magnetization~\cite{bramwellMagnetization1993}; however, attempts to definitively assign a BKT or Ising-type transition have proven challenging owing to the difficulty of accurately measuring the critical exponents in 2D CrSBr via static magnetization and disentangling the role of the sample's thickness in the previous observations~\cite{ziebelCrSBr2024}. 
Notably, evidence consistent with the BKT transition has been reported, but only in bulk samples~\cite{lopez-pazDynamic2022,scheieSpin2022, moroRevealing2022}.
This matter is further complicated by the presence of long-range dipolar
interactions, which are expected to modify the values of the critical exponents~\cite{sunDipolar2024, defenuLongrange2023}.

Ising- and BKT-type criticality can be distinguished based on qualitative differences in the nature of their two-point correlation functions, which manifest themselves in distinct temperature-dependent behaviors of the decoherence rate of a spin qubit probe\cite{SM}.
The BKT transition is marked by an exponential increase in the correlation length as one approaches the critical temperature, and power-law correlations within the ordered phase~\cite{chaikinPrinciples1995}.
As a result, low-frequency fluctuations are present throughout the ordered phase, and the decoherence rate is expected to exhibit a broad feature that persists well below the critical temperature (Fig.~\ref{fig:fig3}b)\cite{SM}.
By contrast, in the Ising-type transition, low-frequency fluctuations are only expected to occur near the critical point where the correlation length of the system becomes comparable with the NV-sample distance $\ell$.
Only the Ising-type case would then predict a sharp peak in the decoherence rate as a function of temperature at the phase transition, which agrees qualitatively with our observations (Fig.~\ref{fig:fig2}e and Fig.~\ref{fig:fig3}b) and  rules out the BKT hypothesis.

While the qualitative features of the characteristic coherence decay rate ($1/TT_2$) can help distinguish between Ising and BKT-type criticality in general, quantitative analysis of the full temperature dependent decoherence decay $C(\tau, T)$ can be used to determine the critical exponents of the phase transition, which reveals deep information about the interactions that govern  criticality in 2D CrSBr. 
In particular, long-range magnetic dipolar interactions have been shown to dictate important features of the propagation of magnons in CrSBr~\cite{sunDipolar2024}, and are also expected to influence magnetization dynamics at criticality.
This suggests that criticality in CrSBr may not be entirely determined by the critical exponents for the conventional short-range two-dimensional Ising universality class~\cite{defenuLongrange2023}.
Understanding the role of long-range interactions and their impact on the critical exponents of the transition requires an analysis of the \emph{functional form} of the measured coherence decay $C(\tau, T)$, which encodes  details about the two-point correlation function via the frequency and momentum dependence of the material's fluctuations~\cite{machadoQuantum2023}.

The connection between the functional form of the decoherence dynamics and the critical properties of the material is most conveniently captured using the language of momentum and frequency filter functions as related to noise spectroscopy \cite{machadoQuantum2023}.
Fluctuations in the CrSBr sample with frequency $\omega$ and wavenumber $q$ (encoded in the structure factor $S(\omega,q)$\cite{machadoQuantum2023,hohenbergTheory1977}) generate fluctuating magnetic fields near the NV probe.
The strength of the magnetic field fluctuations sensed by the probe depends on the wavenumber $q$ and the distance $\ell$ between the two-dimensional sample and the NV sensor; this is captured by the momentum filter function $W_\ell(q)\propto q^3 e^{-2q\ell}$ which is peaked around $q\ell = 3/2$~\cite{machadoQuantum2023}.
By contrast, the frequency response of the probe is determined by the applied pulse sequence which dictates the frequency dependence of decoupling between the NV sensor and its environment; this can be expressed in terms of a frequency filter function $W_{\tau}(\omega)$.
For the spin echo pulse sequence already discussed, $W_{\tau}(\omega)\propto \sin^4(\omega\tau/4)/\omega^2$, encoding the static-noise resilience [$W_\tau(\omega=0) =0$] and maximal response at frequencies $\omega \propto 1/\tau$.
Altogether, one obtains a direct relationship between the decoherence dynamics of the probe at depth $\ell$ (encoded in $\langle \phi^2\rangle$) and the fluctuations in the material:
\begin{equation}\label{eq:phi}
\langle\phi^2(\tau)\rangle =\int_{-\infty}^{\infty}\,\frac{d\omega}{2\pi}W_\tau(\omega) \int_0^\infty \frac{dq}{2\pi} W_\ell(q) \ S(\omega,q)~.
\end{equation}
Integrating the coherence signal over the depth distribution of NVs $\rho(\ell)$ yields the decoherence dynamics of the ensemble (see SM Sec.~2.2).

Near the phase transition, the material's structure factor $S(\omega,q)$ attains a universal form parametrized by the critical exponents associated with the transition ($\nu$, $z$) and both the strength of interactions $J$ and dissipation $\omega_0$ in the material: $S(\omega, q) = S(\omega, q \ |\  \nu, z, \omega_0, J)$\cite{SM, hohenbergTheory1977}.
Taking into account the depth distribution of our NV ensemble and the physical properties of CrSBr (which specify $J$) \cite{ziebelCrSBr2024, SM}, we can simulate the observed decoherence signal and directly compare it with the experimental observations.
Crucially, to isolate the decoherence signal due to critical fluctuations from any other sources of magnetic noise from the sample, we normalize the decoherence dynamics near the critical temperature by the decoherence profile at a temperature well below criticality ($T_{\text{norm}}=130~\mathrm{K}$), which removes signal due to background decoherence mechanisms in the NV ensemble~\cite{SM, deer}. By globally fitting the temperature-normalized decoherence ($-2\log C(\tau)$) across different temperatures $T$ to the simulated ensemble decoherence using a structure factor parametrized by the critical exponents $z$ and $\nu$ and dissipation frequency $\omega_0$ (Fig.~\ref{fig:fig3}c), we quantitatively extract the critical exponents associated with the phase transition \cite{SM}. 
Notably, the goodness of fit (computed via the reduced $\chi^2$) demonstrates that the extracted critical exponents are markedly distinct from the values of the short-range Ising universality class $z_{2D}=2$ and $\nu_{2D}=1$ (Fig.~\ref{fig:fig3}d).
This is a robust feature of our analysis, independent of the value of the dissipation ($\omega_0$), uncertainty in the NV implantation depth distribution, or the other choices in our model and analysis \cite{SM}. 
Indeed, the reduction of both $\nu$ and $z$ from the short-range Ising model is consistent with expectations based on the effects of long-range magnetic dipolar interactions \cite{defenuLongrange2023}; in CrSBr, these arise from magnetic dipolar coupling, whose anisotropic, long-range character is known to shape the material's magnetism~\cite{sunDipolar2024}.

We conclude by noting that, while our experiment can extract both $z$ and $\nu$ critical exponents, the $\chi^2$ highlights the presence of a strong correlation between the extracted $z$ and $\nu$ values (Fig.~\ref{fig:fig3}d).
This is to be expected; our fitting procedure explores the decoherence dynamics only as a function of one parameter -- the temperature of the system -- in an NV ensemble with a fixed depth distribution.
Studying the decoherence dynamics as a function of additional tunable parameters, particularly a tunable NV-sample distance $\ell$, would further constrain the fit and enable a more precise measurement of both $z$ and $\nu$ \cite{machadoQuantum2023}.
Such measurements could be possible by employing tip-based single-NV sensing techniques ~\cite{ simonFiltering2022, ariyaratneNanoscale2018}, by using tunable magnetic field gradients to selectively address NV centers at different depths in our ensemble~\cite{araiFourier2015}, or by leveraging delta-doped samples~\cite{davisProbing2023} to obtain precise, narrow ensemble depth distributions.

\section*{Outlook}
Our work constitutes a first foray into the quantitative analysis of critical phenomena in 2D materials using NV $T_2$ noise magnetometry.
Using this technique, we are able to measure the critical exponents of the magnetization transition for 2D CrSBr and disambiguate its nature: it falls within an Ising universality class modified by the presence of long-range interactions~\cite{defenuLongrange2023}.
Beyond this result, the generality of our approach opens up several applications and directions for further exploration. 

First, achieving further constraints on the values of critical exponents extracted from this technique encourages the development and implementation of experimental platforms able to address NV probes at precise and tunable distance offsets from the sample.
Second, new features can be extracted by exploring different measurement schemes and modalities.
For example, noise covariance magnetometry enables the direct measurement of correlations across two distinct single NVs, avoiding the need to engineer the \textit{q}-space filter function using the NV implantation depth~\cite{rovnyNanoscale2022, rovnyMultiqubit2025}.
Similarly, higher-order contributions to the decoherence dynamics can be extracted by exploring different pulse sequences that target the same frequency window~\cite{sungNonGaussian2019, curtisNongaussian2025}, opening a new regime for exploring intricate correlations accessible by only a few other techniques~\cite{glereanUltrafast2025}.
Finally, while demonstrated in the context of magnetic insulators, the tools explored in this work can be naturally extended to a broad range of materials, phases of matter, and critical phenomena, laying out a path towards the exploration of correlated phenomena in two dimensions.

\bibliography{scibib, notes}

\begin{thebibliography}{10}
\bibitem{t2}
{We define $\tau=T_2$ as the time at which the coherence signal has decayed to a fixed value of $C(T_2) = e^{-2}$, allowing for a fair comparison of the the decoherence rate ($1/T_2$) across coherence decays fit by stretched exponentials.}
\end{thebibliography}

\bibliographystyle{Science}

\section*{Acknowledgments}
We thank Shaowen Chen, Nikola Maksimovic, Ruolan Xue, Laurens Vanderstraeten, Nick Bronn, Jennifer Schloss, Lin Pham, Mikael Backlund and Kristina Liu for fruitful discussions. We thank Daybis Tencio for obtaining the \textsc{C-TRIM} software and assisting in aspects of the sample preparation. NV magnetometry experiments and instrument development at Columbia were primarily supported by Programmable Quantum Materials, an Energy Frontier Research Center funded by the U.S. Department of Energy (DOE), Office of Science, Basic Energy Sciences (BES) under award no. DE-SC0019443. M.E.Z. acknowledges funding support from Columbia University’s Research Initiatives in Science and Engineering program. J.S.O. was supported by the National Science Foundation (NSF) under CHE-2004008. A.N.P. was supported by the Air Force Office of Scientific Research under grant FA9550-21-1-0378. A.B.T. was supported by the NSF under grant DMR-2004691. Work at Princeton was supported by the NSF (QuSEC-TAQS OSI 2326767 and Princeton University’s Materials Research Science and Engineering Center DMR-2011750) and the Gordon and Betty Moore Foundation (grant DOI 10.37807/gbmf12237). T.D. acknowledges support by the DOE Office of Science, National Quantum Information Science Research Centers, Co-design Center for Quantum Advantage (C2QA) under contract number DE-SC0012704. A.L. and C.A.M. acknowledge support from the National Science Foundation, grant 1914945. C.A.M. also acknowledges support from the National Science Foundation, grant 2203904. F.M. acknowledges support from the NSF through a grant for ITAMP at Harvard University, and the Netherlands Organisation for Scientific Research (NWO/OCW), as part of Quantum Limits
(project number SUMMIT.1.1016). S.C. acknowledges support from a PQI Community Collaboration Award.

\section*{Author Contributions}
M.E.Z. conceived of the project with A.N.P., P.J.S., and J.S.O. M.E.Z. performed the experiments. M.E.Z. and F.M. analyzed the data with E.J.D. and S.C. M.E.Z.  built the experimental apparatus with assistance from B.U. and designed the experiments with A.L., T.D., and C.A.M.  Z.Y. and N.P.dL. processed the diamond used for ensemble NV magnetometry and A.B.T. fabricated the CrSBr flake on the diamond. D.G.C., E.T., Mi.E.Z., and X.R. synthesized and provided CrSBr crystals. N.Z. and K.L.S. assisted with testing the microwave PCB. D.W.D. grew and provided NV ensemble samples for initial experiments to verify the setup. X-Y.Z. and J.C.H. provided equipment and resources for experiments. A.N.P., P.J.S., and J.S.O. supervised the work. M.E.Z., F.M., E.J.D., and S.C. wrote the  manuscript. The manuscript reflects the contributions and ideas of all authors.

\section*{Supplementary Materials}
Materials and Methods\\
Supplementary Text\\
Figs. S1 to S16\\
References \cite{sangtawesinOrigins2019, alsidPhotoluminescence2019, posseltComputer1992, raatzInvestigation2019, bauchDecoherence2020, wangOneDimensional2013, bucherQuantum2019, durhamScanning2018, aitchisonLumpedcircuit1971, guptaComputerAided1981, rabeauImplantation2006, vanderwaltScikitimage2014, ortnerMagpylib2020, dreauAvoiding2011, schweigerPrinciples2001, puzzuoliQiskit2023, barrySensitivity2020}

\clearpage

\clearpage

\clearpage
\begin{figure*}
\centering
\includegraphics[width=0.8\linewidth]{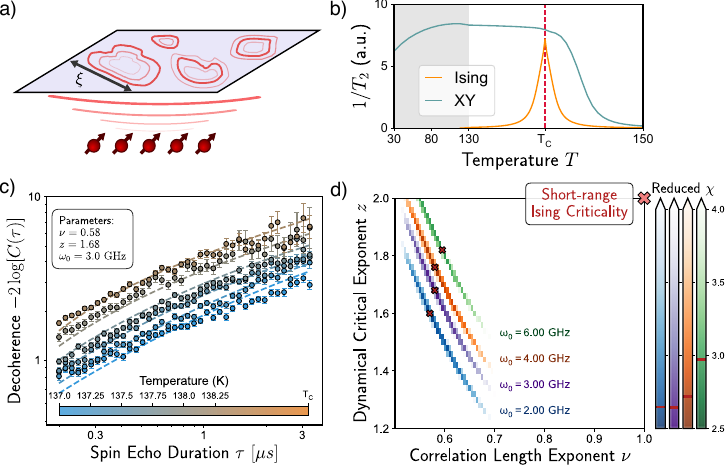}
\caption{\textbf{Analysis of 2D critical dynamics in CrSBr.}~
{\bf (a)} Near criticality, the correlation length $\xi$ of spin fluctuations in CrSBr will approach the spin-qubit distance $\ell$, which sets the wave vector $q$ of probed fluctuations as $q\sim \ell^{-1}$.
At this point, the noise experienced by the NV is maximal.
{\bf (b)} The presence of quasi-long-range order deep in the XY ferromagnetic phase (associated with a BKT transition) leads to a broad temperature range where low-frequency fluctuations strongly enhance the NV decoherence rate ($1/T_2$).
By contrast, in an Ising magnet, such low-frequency fluctuations are only present near the phase transition, leading to a single sharp feature at the phase transition temperature.
{\bf (c)} Global fit to the temperature dependent NV spin-echo decoherence ($\langle\phi^2(\tau)\rangle$) near criticality using the universal scaling form of the structure factor $S(\omega, q)$ that describes spin fluctuations in the material.
Informed by the details of CrSBr, the structure factor is parametrized only by the critical exponents $\nu$ and $z$ and a dissipative rate $\omega_0$. 
The overall best fit parameters are $\nu=0.58$, $z=1.68$, and $\omega_0=3.0$~GHz. 
{\bf (d)} Landscape of reduced $\chi^2$ as a function of  $\nu$ and $z$ for different choices of $\omega_0$. 
Despite a strong correlation between $\nu$ and $z$ across different $\omega_0$, the short-range Ising critical exponents (top right corner) are inconsistent with our observations.     
}
\label{fig:fig3}

\end{figure*}

\clearpage
\supplementarysection

\begin{center}
{\huge \textbf{Supplementary Materials for:}} 
\\~\\
{\LARGE Quantum Noise Spectroscopy of Criticality in an Atomically Thin Magnet}
\\~\\
Mark E. Ziffer$^{1,2,3,\P,\dagger,*}$, Francisco Machado$^{4,5,6,\dagger,*}$, Benedikt Ursprung$^2$, \\Artur Lozovoi$^{7,8}$, Aya Batoul Tazi$^1$, Zhiyang Yuan$^7$, Michael E. Ziebel$^3$,\\ Tom Delord$^8$, Nanyu Zeng$^9$, Evan Telford$^3$, Daniel G. Chica$^3$,\\
Dane W. deQuilettes$^{10,11}$, Xiaoyang Zhu$^3$, James C. Hone$^2$, Kenneth L. Shepard$^9$, 
\\Xavier Roy$^3$, Nathalie P. de Leon$^7$, Emily J. Davis$^{12}$, Carlos A. Meriles$^{8,*}$,\\ Jonathan S. Owen$^{3,*}$, P. James Schuck$^{2,*}$, Shubhayu Chatterjee$^{13,*}$,\\
Abhay N. Pasupathy$^{1, 14,*}$
\\~\\

\small{$^{1}$Department of Physics, Columbia University, New York, NY 10027 USA}\\
\small{$^2$ Department of Mechanical Engineering, Columbia University, New York, NY 10027 USA}\\
\small{$^3$ Department of Chemistry, Columbia University, New York, NY 10027 USA}\\
\small{$^4$ ITAMP, Harvard-Smithsonian Center for Astrophysics, Cambridge, MA 02138, USA}\\
\small{$^5$ Department of Physics, Harvard University, Cambridge, MA 02138, USA}\\
\normalsize{$^6$ QuTech and Kavli Institute of Nanoscience, Delft University of Technology,}\\
\normalsize{Delft, 2600 GA, The Netherlands}\\
 \small{$^7$ Department of Electrical and Computer Engineering, Princeton University, Princeton, NJ 08540, USA}\\
\small{$^8$ Department of Physics, CUNY, The City College of New York, New York, NY, 10031, USA}\\
\small{$^9$ Department of Electrical Engineering, Columbia University, New York, NY 10027 USA}\\
\small{$^{10}$Lincoln Laboratory, MIT, Lexington, MA, 02421, USA}\\
\small{$^{11}$Center for Quantum Engineering, MIT, Cambridge, MA, 02139, USA}\\
\small{$^{12}$Department of Physics, New York University, New York, NY 10003, USA}\\ 
\small{$^{13}$Department of Physics, Carnegie Mellon University, Pittsburgh, PA 15213, USA}\\
\small{$^{14}$Condensed Matter Physics and Materials Science Department, Brookhaven National Laboratory, Upton, NY, 11973 USA}
\\~\\
\normalsize{†These authors contributed equally}
\\~\\
\normalsize{$^\ast$Corresponding author. Email:   apn2108@columbia.edu, p.j.schuck@columbia.edu,}\\ \normalsize{jso2115@columbia.edu, mark@qunett.com,
cmeriles@ccny.cuny.edu,}\\ \normalsize{shubhayuchatterjee@cmu.edu}, \normalsize{f.machado@tudelft.nl}
\\~\\
\normalsize{$\P$ Present address: Quantum Network Technologies, Inc., Boston, MA 02215}

\end{center}
\clearpage

\section{Materials and Methods}

\subsection{NV Ensemble Preparation and Characterization}
\label{sec:SM_samplePrepAndCharacterization}

\begin{figure}
    \centering
    \includegraphics[width=1\linewidth]{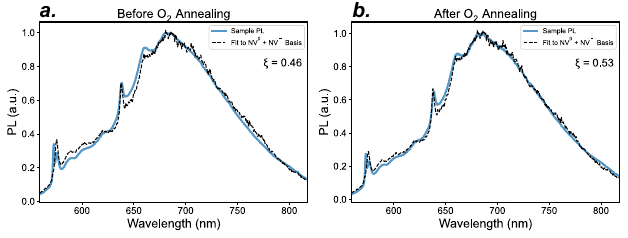}
    \caption{\textbf{Photoluminescence characterization of the NV ensemble charge state ratio.}~We determined the NV$^-$ to NV$^0$ ratio before and after the O$_2$ annealing steps of our sample preparation using the photoluminescence decomposition method of Alsid \textit{et al.} \cite{alsidPhotoluminescence2019}. Photoluminescence spectra were measured on a confocal fluorescence set-up using $\sim1.8 ~\mathrm{mW}$ 532 nm CW laser excitation focused onto the sample with a 0.6 NA objective (Olympus LUCPlanFL N, 40x). Dichroic and longpass filters were used to filter the laser excitation and the fluorescence spectra were recorded with a spectrograph (Acton SP2300i) and CCD camera (Princeton Instruments Pixis 400BRX). The data were intensity corrected by measuring the spectrum of a calibrated tungsten halogen light source (StellarNet SL1-CAL). The fraction of the total NV yield in the NV$^-$ charge state ($\xi$) was determined by fitting the spectra to a sum of basis functions $c_-$ and $c_0$  (corresponding to the emission spectra of NV$^-$ and NV$^0$, respectively), and weighting $c_0$ by a factor $\kappa_{532}$ corresponding to the difference in emission rate between the NV$^-$ and NV$^0$ charge states excited at 532 nm, using the equation: $\xi = \frac{[NV^-]}{[NV^-] + [NV^0]} = \frac{c_-}{c_- + \kappa_{532}c_0}$ \cite{alsidPhotoluminescence2019}. O$_2$ annealing was found to improve the total NV$^-$ fraction by $\sim15\%$ ($\xi = 0.46$ before O$_2$ annealing to $\xi = 0.53$ after O$_2$ annealing).} 
\label{fig:SM_NVChargeState}
\end{figure}

A $[100]$-oriented electronic grade diamond substrate with a 
$\sim2~\mathrm{\mu m}$ thick isotopically pure $^{12}\mathrm{C}$ overgrown 
diamond layer was purchased from QuantumDiamonds, GmbH. The uncertainty in the miscut angle of the surface facet orientation was $\pm 3\degree$. The as-received diamond was cleaned by ultrasonication followed by tri-acid cleaning in a boiling mixture of 1:1:1 perchloric:nitric:sulfuric acid refluxing for 2 hours. The 
$^{12}$C surface was then implanted with $^{15}$N$^+$ ions with $25\:$keV implantation energy, $2\times10^{12}\:$cm$^{-2}$ dose, and $5\degree$ sample tilt (Coherent, Inc., Innovion Corp., San Jose, CA). After ion implantation, the sample was cleaned with the tri-acid procedure and evaluated for surface contamination using XPS. The diamond was then annealed under vacuum ($<1\times10^{-5} \:$Torr) using the 
ramped annealing procedure reported by Sangtawesin \textit{et al.} \cite{sangtawesinOrigins2019} without the $1200\degree$ C step, followed by tri-acid cleaning. The sample was then annealed under O$_2$ gas at $455\degree$ C following the procedure in Sangtawesin \textit{et al.}\cite{sangtawesinOrigins2019} in order to form an oxygen terminated surface to stabilize the NV$^-$ charge state for NVs formed at shallow depths within the implanted $^{15}$N distribution (see Figure \ref{fig:SM_NVChargeState} for NV charge state characterization). Following O$_2$ annealing the diamond was cleaned with a 1:2 hydrogen peroxide:sulfuric acid piranha solution. In between steps the diamond was sonicated with Milli-Q water followed by sonication in electronic grade 2-propanol (Sigma Aldrich, 733458). The depth distribution of implanted $^{15}$N ions was determined from \textsc{C-TRIM} simulations as described below.

\subsubsection{C-TRIM Depth Distribution Simulation}
\label{sec:SM_CTRIMsim}

\begin{figure}
\centering
\includegraphics[width=0.7\textwidth]{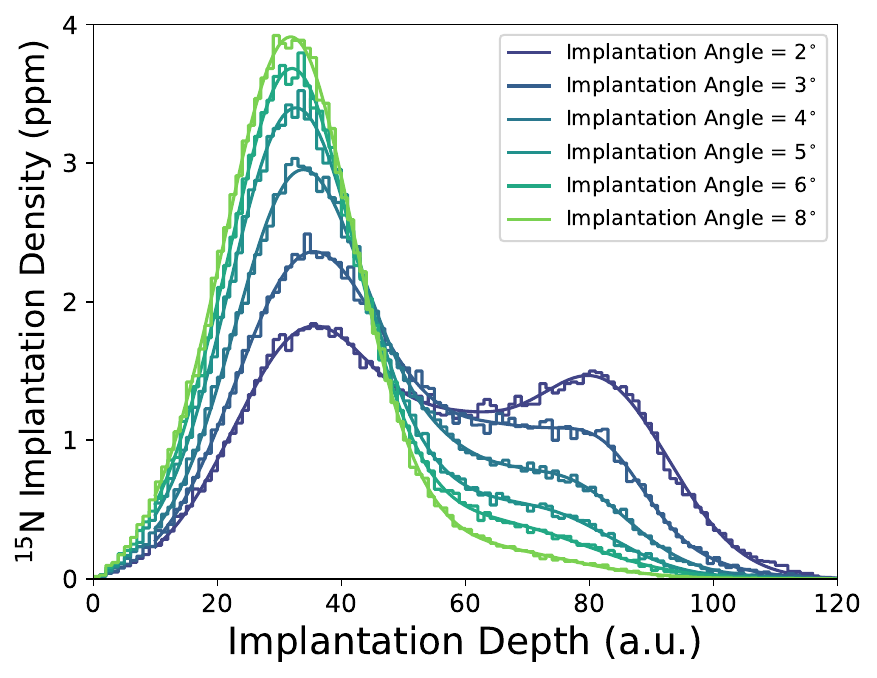}
\caption{\textbf{C-TRIM $^{15}\mathrm{N}$ ion implantation simulation.} C-TRIM simulation of the $^{15}\mathrm{N}$ implantation density as a function of depth (see Section~\ref{sec:SM_CTRIMsim} for simulation details) and implantation angle in degrees (colors).}
\label{fig:SM_NVDensity}
\end{figure}

We modeled the $^{15}\mathrm{N}$ (and NV) depth distribution by performing \textsc{C-TRIM}\cite{posseltComputer1992, raatzInvestigation2019} simulations for $25$~keV $^{15}$N$^+$ ion implantation into $[100]$ oriented diamond. \textsc{C-TRIM} was used (as opposed to \textsc{SRIM}) to  model the tail of the ion distribution at larger depths due to ion channeling \cite{raatzInvestigation2019}. Based on the parameters determined by Raatz \textit{et al.}\cite{raatzInvestigation2019} we used a value of $C_{el}=1.715$ for an implantation energy of $25$ keV. We additionally simulated surface roughness by adding a $1$~nm thick amorphous carbon layer with the density of diamond ($3.51$ g/cm$^3$)\cite{raatzInvestigation2019}. Simulations were performed over a range of tilt angles spanning $5 \pm 3\degree$ to analyze variations in the implantation profile within the range of uncertainty over the miscut angle of our diamond sample (Section \ref{sec:SM_samplePrepAndCharacterization}).  To model an estimated probability density function for the NV depth $\ell$ from the C-TRIM simulations, we fit the simulated $^{15}$N implantation density (Figure \ref{fig:SM_NVDensity}, left axis) to a sum of three truncated Gaussians, which was then normalized to an area of 1 to model an estimated probability density function $\rho(\ell)$ (right axis, Figure \ref{fig:SM_NVDensity}). Note that the analysis in Figure \ref{fig:fig3} of the main text uses $\rho(\ell)$ corresponding to an implantation angle of $5\degree$ (the nominal implantation angle assuming $0 \degree$ sample miscut). We also performed the same analysis over the whole range of possible variation in implantation angle due to sample miscut (Fig.~\ref{fig:SM_ImplantationAngle}) and found that the conclusions regarding the critical exponents do not depend on uncertainty in the implantation angle (see Section~\ref{sec:SI_EnsembleCoherece} for more details).  

\subsubsection{Analysis of the Intrinsic NV Coherence Time}
\label{sec:SM_IntrinsicNVCoherence}

\begin{figure}
    \centering
    \includegraphics[width=1\linewidth]{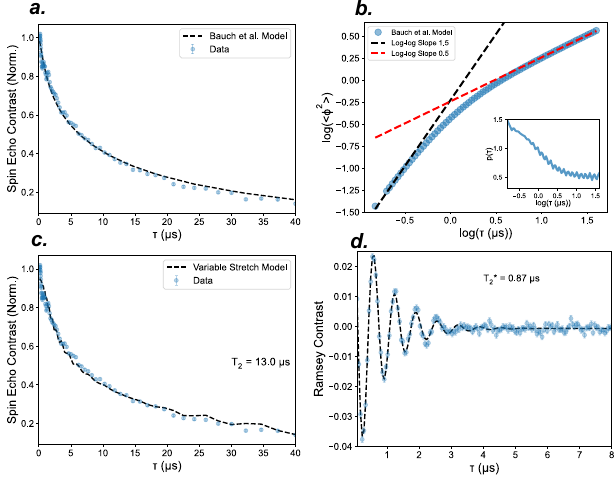}
    \caption{\textbf{Spin coherence of the NV ensemble.}~(a) Spin echo coherence decay of the NV ensemble at $137.5 \:K$ fit to the model of Bauch \textit{et al.}\cite{bauchDecoherence2020} for ensemble averaged NV interactions with the bulk diamond spin bath (see Section \ref{sec:SM_IntrinsicNVCoherence}). (b) Power law scaling of $\langle\phi(\tau)^2\rangle$ determined by linear fits to $\log(\langle\phi(\tau)^2\rangle)$ vs $\log(\tau)$ derived from the model in (a). Inset shows the $\tau$ dependence of the exponential stretch parameter $p(\tau)$ of the coherence decay, determined by calculating $d\log(\langle\phi(\tau)^2\rangle) / d\log(\tau)$. (c) Fit of the data in (a) to $e^{-(\frac{\tau}{T_2})^{p(\tau)}}$ using $p(\tau)$ determined in (b). (d) Free induction decay of the NV ensemble measured by a Ramsey pulse sequence at $137.5 \:K$ (see Figure \ref{fig:SM_PulseSequencePlots}).}
\label{fig:SM_EnsembleCoherence}
\end{figure}

To characterize the intrinsic spin coherence properties of the NV ensemble in our diamond sample, we measured  $T_2$ using spin echo  and  $T_2^*$ using Ramsey at a position $\sim1 \: \mathrm{\mu m}$ off of the flake at $\sim137.5\: $K.  To compare the results with models for NV ensemble coherence, we first fit the spin echo coherence decay by numerically solving the ensemble averaged coherence model from Bauch \textit{et al.}\cite{bauchDecoherence2020}. Here, the coherence decay for a single NV undergoing decoherence from the diamond spin bath is defined as $C(\tau)=e^{-\chi_{\text{SE}}(\tau)}$, with $\chi_{\text{SE}}(\tau) = \Delta_{\text{single}}^2\tau_c^2[\frac{\tau}{\tau_c}-3-e^{-\tau/\tau_c}+4e^{-\tau/2\tau_c}]$ where $\Delta_{\text{single}}^2$ determines the coupling strength to fluctuations from the diamond spin bath which have a characteristic fluctuation correlation time $\tau_c$\cite{bauchDecoherence2020}. The coherence for an NV ensemble is obtained by solving $C_{\text{ens}}(\tau)=\iint e^{-\chi_{\text{SE}}(\tau)}P(\Delta_{\text{single}})P(\tau_c)\,d\Delta_{\text{single}}\, d\tau_c $ where $P(\Delta_{\text{single}})$ is a probability distribution function for $\Delta_{\text{single}}$ characterized by a shape parameter $\Delta_{\text{ens}}$ and $P(\tau_c)$ is a probability distribution for $\tau_c$ with shape parameters $\tau_{c,ens}$  and $\lambda$ \cite{bauchDecoherence2020}.  
Figure \ref{fig:SM_EnsembleCoherence}(a) shows the measured spin echo coherence $C(\tau)$ and  best fit to the model of Bauch \textit{et al.} with $\Delta_{\text{ens}}=0.4\:$MHz, $\tau_{c,\text{ens}}=0.5\:\mathrm{\mu s}$, and $\lambda=3.57$. 
Defining $C(\tau)=e^{-\langle\phi^2(\tau)\rangle/2}$ as in the main text, we then evaluated the power law behavior of $\langle\phi^2(\tau)\rangle$ from the model for $C_{\text{ens}}(\tau)$ by plotting  $\langle\phi^2\rangle$ vs $\tau$  on a log-log scale (Figure \ref{fig:SM_EnsembleCoherence}(b)). 
Linear fitting of the early and late time log-log behavior of $\langle\phi^2\rangle$ vs $\tau$  recovers the  expected  $3/2$ and $1/2$ power law behavior for $\tau<\tau_{c,\text{ens}}$ and $\tau>\tau_{c,\text{ens}}$ respectively for a 3D spin bath (Figure \ref{fig:SM_EnsembleCoherence}(b)) \cite{davisProbing2023}. 
To determine $T_2$ we fit the coherence decay to a function $C(\tau) = e^{-(\frac{\tau}{T_2})^{p(\tau)}}$ where we input a $\tau$ dependent variation in the stretch parameter $p(\tau)$ which is determined by calculating the derivative of $\log(\langle\phi^2\rangle)$ vs $\log(\tau)$ from (Figure \ref{fig:SM_EnsembleCoherence}(b)). 
The calculated $p(\tau)$ is shown in the inset of Figure \ref{fig:SM_EnsembleCoherence}(b). 
Figure \ref{fig:SM_EnsembleCoherence}(c) shows a fit of the data to $e^{-(\frac{\tau}{T_2})^{p(\tau)}}$, from which we determine $T_2 = 13.0 \:\mathrm{\mu s}$. 
Based on the results of Bauch \textit{et al.} \cite{bauchDecoherence2020} we would expect $T_2$ in the range of 10's of $\mathrm{\mu s}$ for a bulk NV ensemble formed from 2.5 ppm nitrogen density. We suspect that due to the spread of NV depths near the surface in our ensemble (Fig \ref{fig:SM_NVDensity}) there is likely additional interaction between the NV ensemble and a surface spin bath (with possible additional contributions of surface spin noise due to residual polymer from the CrSBr transfer), which reduces $T_2$.
We also measured $T_2^*$ using a $\frac{\pi}{2}_{(x)}\rightarrow \frac{\pi}{2}_{(x, -x)}$ Ramsey pulse sequence (see Figure \ref{fig:SM_PulseSequencePlots} for pulse sequence). Figure \ref{fig:SM_EnsembleCoherence}(d) shows a fit of the Ramsey  data to a function of the form  $C_{\text{Ramsey}}(\tau)\propto e^{-(\tau/T_2^*)}\cos{(\omega_R \tau)}$, where $\omega_R$ is the Ramsey interference frequency. From the fits to the Ramsey data we obtain $T_2^*=0.87\:\mathrm{\mu s}$.

\subsection{CrSBr Exfoliation and Transfer}
\label{sec:SM_CrSBrPrep}

 Bulk crystals of CrSBr were synthesized using a chemical vapor transport reaction with a temperature gradient of $950\degree$~C to $850\degree$~C as described in detail in reference \cite{scheieSpin2022}. Few-layer flakes of CrSBr were prepared by Scotch tape exfoliation from a bulk CrSBr crystal onto SiO$_2$(285 nm)/Si wafers. The thickness of the flakes was determined using atomic force microscopy and correlated optical contrast \cite{telfordCoupling2022}. Transfer of CrSBr flakes onto the diamond substrate was performed using a standard dry stamp method \cite{wangOneDimensional2013} with polycarbonate (PC) as the pick-up polymer. The substrate was oriented on a rotation stage such that the long axis of the CrSBr flake of interest was aligned roughly parallel with a $[110]$ edge of the diamond substrate as seen under the microscope camera prior to transfer. Following transfer the PC was washed off from the diamond with chloroform. 

\subsection{Confocal ODMR Setup}
\label{sec:SM_ODMRSetup}

\begin{figure}
    \centering
    \includegraphics[width=1\linewidth]{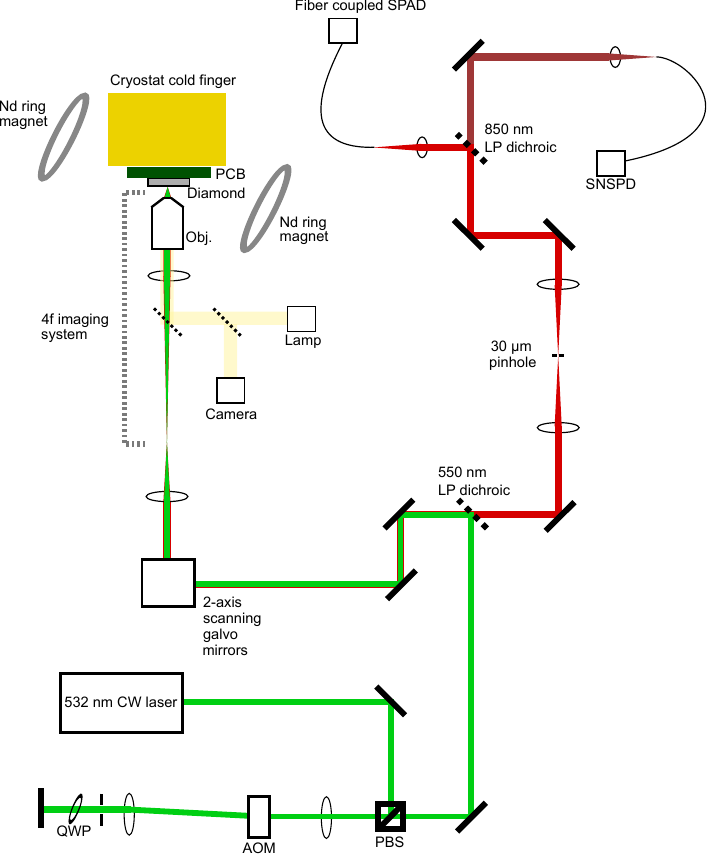}
    \caption{\textbf{Experimental set-up.}~Diagram of the confocal ODMR setup used in this work.}
\label{fig:SM_ConfocalSetup} 
\end{figure}

\begin{figure}
    \centering
    \includegraphics[width=0.75\linewidth]{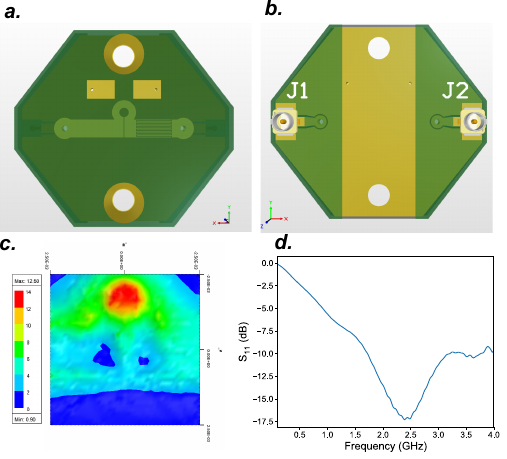}
    \caption{\textbf{PCB for microwave delivery.}~For microwave delivery we designed a planar 50-Ohm transmission-line microstrip LC resonator consisting of an $\Omega$-loop planar inductor in series with an interdigitated planar capacitor\cite{aitchisonLumpedcircuit1971, guptaComputerAided1981} fabricated onto a PCB (top view in (a), bottom view in (b)) . The transmission line is terminated with a 50 Ohm resistor connected to J1 (shown in (b)) via a 50 Ohm coax cable, such that the impedance at the LC resonance frequency is 50 Ohms. The microstrip width and dimensions of the planar $\Omega$-loop inductor and interdigitated capacitor were determined by optimizing the simulated $S_{11}$ in HFSS (Ansys) to obtain a broad LC resonance centered around $\sim2.7 \:$GHz matched to an input impedance of 50 Ohms.   In (c) we show an HFSS simulation of the H-field near resonance at the surface of a 0.3 mm thick diamond placed on top of the PCB, where we achieve a $\sim0.5~ \mathrm{mm^2}$  area of uniform maximum H-field above the center of the $\Omega$-loop. In cryogenic experiments, an area of the ground plane not covered with solder mask (large gold rectangular area at the center of the PCB in (b)) is mounted directly onto the cold finger of the cryostat contacted with silver paste. The ground plane is connected to two cooling pads through vias (gold rectangular areas above the $\Omega$-loop in (a)), which contact the diamond mounted on the top side of the PCB. In (d) we show $S_{11}$ measured with a VNA (Agilent N5230A / N4431-60003) on a 50-Ohm terminated PCB manufactured from our design, which shows a broad LC resonance near $2.5 \:$GHz. }
\label{fig:SM_OmegaLoop}
\end{figure}

All experiments were performed on a home-built laser scanning confocal optically detected magnetic resonance (ODMR) microscope built around a Janis ST-500 optical cryostat (Figure \ref{fig:SM_ConfocalSetup}). A 532 nm laser (Coherent Sapphire SF 532-50) is sent through an acousto-optic modulator (Isomet M1205-P80L-1) in a double-pass configuration for pulse modulation and directed via a 550 nm  longpass dichroic mirror (Thorlabs DMLP550R) to a set of galvo-mirrors (Thorlabs GVS002) positioned in the back focal plane of the scan lens (Thorlabs SL50-CLS2) in a $4f$ imaging system, where a tube lens (Thorlabs TTL200-S8) and a microscope objective (Nikon Plan Fluor 60x, 0.7NA) image the scanning laser spot onto the sample. Typical CW laser power on the sample is $\sim 1\:$mW. Photoluminescence (PL) from the sample is spectrally filtered with a 550 nm longpass dichroic mirror and additional 650 nm longpass filters, after which it is  spatially filtered with a $30\: \mathrm{\mu m}$ pinhole. The PL is then split by a 850 nm dichroic mirror to separate NV PL (in the range of 650-850 nm) from the CrSBr PL at $\sim 930$ nm. The NV PL is sent to a fiber coupled single photon counting APD (Excelitas SPCM-AQRH-15-FC), and the CrSBr PL is sent to a superconducting nanowire single photon detector (Single Quantum EOS-6). Photon counts from the detectors are read out using gated counter channels on a DAQ (NI USB-6343). Microwave delivery was achieved by mounting the diamond onto a PCB resonator, where an $\Omega$-loop planar inductor delivered homogeneous RF magnetic fields over an area of around $\sim 0.5 \:\mathrm{mm}^2$  (see Figure \ref{fig:SM_OmegaLoop} for details). The diamond was positioned on the PCB under a microscope such that the CrSBr flake was near the center of the $\Omega$-loop. The diamond was silver pasted onto the PCB and positioned such that it contacted cooling pads connected by vias to the ground plane of the PCB which is in thermal contact with the cold finger of the cryostat (see Figure \ref{fig:SM_OmegaLoop}). Microwaves are sourced from a signal generator (Stanford Research Systems SG384/3) and passed through a gated  RF switch (MiniCircuits ZASWA-2-50DRA+) to generate pulses. Microwave pulses are then sent through an amplifier (MiniCircuits ZHL-16W-43-S+) and sent to the PCB. The output of the amplifier is protected with a circulator (DiTom D3C2040). Phase modulation of the microwave pulses is achieved by sending input signals (generated by gating a DC voltage with MiniCircuits ZASWA-2-50DRA+ switches) to the IQ modulator built into the SG384/3 \cite{bucherQuantum2019}. A PulseBlaster ESR-Pro-500 pulse generator is programmed to deliver TTL pulses through output channels to  control modulation of the AOM, gating of microwave switches, and gating of DAQ counter channels (see Section \ref{sec:SM_PulseSequences} and Figure \ref{fig:SM_PulseSequencePlots}). Two 3" OD ring magnets (K\&J Magnetics RZ0Y0X0) are mounted in a coaxial configuration outside the cryostat on a 5-axis stage and are adjusted such that a  $\sim3.5 \:$mT  bias field is aligned along a $[111]$ axis of the diamond (see Section \ref{sec:SM_Magnetometry} and Figure \ref{fig:SM_NVMagDetails}). The cryostat is cooled with a closed-cycle Helium recirculating system (Janis RGC4) and the temperature is maintained using a PID temperature controller (Lakeshore 335). All instruments and experiments are controlled using custom-written Python code with the ScopeFoundry platform \cite{durhamScanning2018}.

\subsection{Pulse Sequences and Contrast Definitions}
\label{sec:SM_PulseSequences}

The detailed pulse sequences for pulsed ODMR, spin echo, Rabi, Ramsey, and CW ODMR experiments are shown in Figure \ref{fig:SM_PulseSequencePlots}.
For pulsed ODMR, Rabi and CW ODMR experiments, the raw contrast is defined as $\frac{C_{\text{S}} - C_{\text{R}}}{C_{\text{R}}}$, where $C_{\text{S}}$ and $C_{\text{R}}$ correspond to the``signal'' and ``reference'' data from the readout channels of the DAQ. 
For spin echo (or Ramsey) experiments raw contrast is defined as:  
\begin{equation}\label{eq:contrast}
2\frac{C_{\text{S}}- C_{\text{R}}}{C_{\text{S}} + C_{\text{R}}} 
\end{equation}

Equation \ref{eq:contrast} is appropriate to use in an experiment where the phase ($\phi$) accumulated during a free precession time $\tau$ is measured by mapping $\phi$ onto a population difference of basis states of the observable $\langle S_z\rangle$ (i.e. the $\ket{\uparrow}$ and $\ket{\downarrow}$ basis). Equation \ref{eq:contrast} is equivalent to writing $\frac{n_{\ket{\uparrow}}-n_{\ket{\downarrow}}}{n_{avg}}$, where $n_{\ket{\uparrow}}$ and $n_{\ket{\downarrow}}$ are the average number of photons measured after mapping onto the  $\ket{\uparrow}$ and $\ket{\downarrow}$ basis respectively, and ${n_{avg}}$ is the total averaged photon count \cite{barrySensitivity2020}.

\subsection{Magnetometry}
\label{sec:SM_Magnetometry}

For all AC and DC magnetometry measurements the bias magnetic field is aligned such that the B-field direction is projected along one of the four possible [111] directions of diamond, corresponding to the spin quantization axis of one of four possible subsets of NV orientations in the ensemble \cite{bucherQuantum2019}. Figure \ref{fig:SM_NVMagDetails}(a) shows the CW ODMR spectrum representing a typical magnetic field alignment. The field is aligned so that the ODMR spectra for the three ``off-axis'' NV orientations (orientations which have the magnetic field projected partially off-axis from the spin quantization axis) collapse onto one peak with the $\sim3\:$MHz $^{15}\mathrm{N}$  hyperfine splitting \cite{rabeauImplantation2006} overlapped when measured with low microwave power (inset of Fig \ref{fig:SM_NVMagDetails}(a)). Rabi measurements (see Figure \ref{fig:SM_TempDepOmegaGammaRatio}(d)) are then performed at the  $|{m_s=0}\rangle\rightarrow|{m_s=-1}\rangle$ resonance frequency for the ``on axis" NV orientation to determine the $\pi-$ pulse duration for all subsequent coherent pulsed experiments. For pulsed ODMR imaging experiments, seven frequency points are measured which correspond to zero-crossings of 1st, 2nd, 3rd and 4th derivatives of a Lorentzian fit determined from a spectrum initially measured at a position $\sim1~\mathrm{\mu m}$ away from the flake (Figure \ref{fig:SM_NVMagDetails}(b)). These seven points correspond to the main points of curvature of a Lorentzian function which allows us to fit the pulsed ODMR data measured at each pixel to a full Lorentzian function to create images both of the resonance frequency and pulsed ODMR contrast (see Section \ref{sec:SM_TemperatureNormalization}).
\subsection{Sample Drift Correction}
\label{sec:SM_sampleDrift}

\begin{figure}
    \centering
    \includegraphics[width=0.8\linewidth]{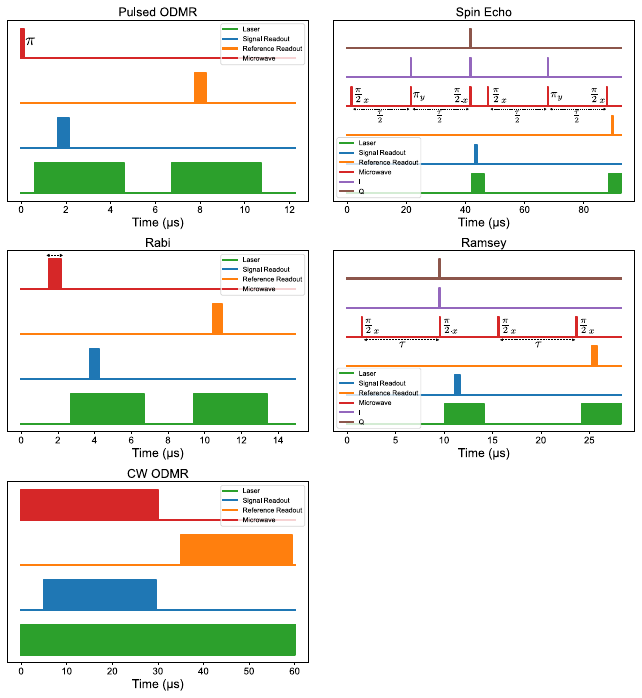}
    \caption{\textbf{Full pulse sequence diagrams.}~Here we show the timing sequences for the TTL pulse outputs from channels of the PulseBlaster which control digital modulation of the AOM (labeled ``Laser''),  gating of the RF switch that feeds the microwave amplifier from the SG384/3 signal generator (labeled ``Microwave"), gating of  DAQ counter channels for signal and reference readout of the APD (labeled ``Signal Readout" and ``Reference Readout"), and gating of RF switches which feed a DC voltage into the I and Q modulation inputs of the SG384/3 signal generator (labeled ``I" and ``Q").  For pulsed ODMR, spin echo, Rabi, and Ramsey, the signal and reference readout pulses are set to $500$ ns and laser pulses are set to a duration of $4$ $\mu$s. A delay between the laser pulse onset and the signal and reference readout pulses is set to compensate for an electronic delay from the instruments \cite{bucherQuantum2019}. The I and Q modulation pulses are padded by $20$ ns with respect to the $\pi$ and $\pi/2$ pulses in spin echo and Ramsey to compensate for the response time of the switches and SG384/3 IQ modulation  \cite{bucherQuantum2019}.}
\label{fig:SM_PulseSequencePlots}
\end{figure}

\begin{figure}
    \centering
    \includegraphics[width=1\linewidth]{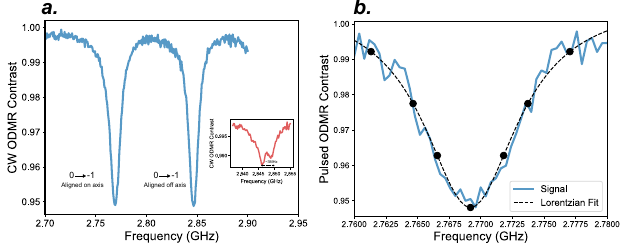}
    \caption{\textbf{NV magnetometry details.}~(a) CW ODMR spectrum corresponding to alignment of the bias magnetic field along a $[111]$ crystallographic axis of diamond. The ODMR peak labeled ``Aligned on axis" corresponds to the $|m_s=0\rangle\rightarrow|m_s=-1\rangle$ resonance of NVs with their spin quantization axis oriented along the same direction as the bias field. (Inset) CW ODMR spectrum of the ``off axis'' $|m_s=0\rangle\rightarrow|m_s-1\rangle$ resonance measured with low microwave power to resolve the overlap of the $^{15}\mathrm{N}$ hyperfine lines of NVs with spin quantization axes corresponding to the three ``off axis" $[111]$ orientations. (b) Pulsed ODMR spectrum of the ``on axis" $|m_s=0\rangle\rightarrow|m_s-1\rangle$ fit to a Lorentzian function, showing the seven measurement points that are selected for pulsed ODMR imaging (see Section \ref{sec:SM_Magnetometry} for details).}
\label{fig:SM_NVMagDetails}
\end{figure}

In pulsed ODMR imaging experiments, the average count rate from the SNSPD (which records PL from CrSBr) is recorded at each pixel along with the data from the pulsed ODMR pulse sequence (Figure \ref{fig:SM_PulseSequencePlots}), such that PL images of the CrSBr flake and pulsed ODMR images can be constructed from data recorded simultaneously. Typically 460 pixels are recorded for each image with 5000 pulsed ODMR averages per pixel (averaging $\sim1.6\:$s/pixel and a total of $\sim 12 \:$min/image), and  $>20$ images are recorded which amounts to a total of $>10^5$ pulsed ODMR averages per pixel. Over the course of a several hour measurement, the image will drift slowly (over 10's of minutes) on the scale of 100's of nm to $\sim \mathrm{\mu m}$,  which is found to be correlated with thermal fluctuations in the room and is likely due to thermal drift of the optics. This drift is corrected in image post-processing by cross correlating the first CrSBr PL image recorded in the measurement sequence with each image in subsequent sequences. The drift is calculated with sub-pixel accuracy using the \textsc{scikit-image} phase cross correlation routine\cite{vanderwaltScikitimage2014} and is determined as a function of time using the image file timestamps. Image interpolation is then used to correct pixel assignments for images recorded at different times in the measurement and reconstruct  drift-corrected PL and pulsed ODMR images.

For spin echo experiments drift correction is performed on the fly so that a fixed measurement position on the sample is maintained over long measurement periods. An initial PL image of the CrSBr flake is recorded and position coordinates on the PL image to perform spin echo measurements are selected. A series of spin echo pulse sequences (typically $10^5$) for 75 different interpulse delay times ($\frac{\tau}{2}$) are then performed at the given coordinates which takes $\sim6-7 \:$min. Another PL image is then measured and cross-correlated with the initial PL image to determine the sample drift and the target coordinates for the next series of spin echo measurements are updated to correct for drift. The process is repeated $75-100$ times for a total of $10^6$ to $10^7$ averages of the spin echo pulse sequence. The typical spread of coordinates sampled over the total course of  spin echo measurements (with variation due to differences in the results of image cross correlation) is shown by the red dots in Figure \ref{fig:SM_SampleDrift}.
\subsubsection{Pixel-to-Pixel Variation in Decoherence Dynamics}
\label{sec:SM_spatialVariation}

\begin{figure}
    \centering
    \includegraphics[width=0.5\linewidth]{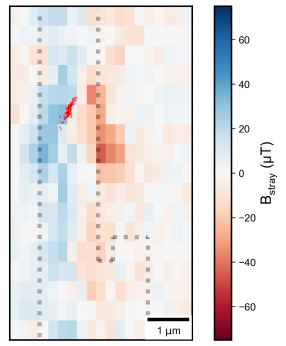}
    
    \caption{\textbf{Spread of points sampled in spin echo measurements with drift correction.}~Pulsed ODMR stray field image recorded at $137.5 \:$K overlaid with the spread of points sampled in spin echo measurements at this temperature.}
\label{fig:SM_SampleDrift}

\end{figure}

Due to slight variations from the image cross-correlation routine used to update the laser position for drift correction, each individual spin-echo pulse sequence repetition samples a slightly different area around the target position on the CrSBr flake, which is logged and tracked (see Figure \ref{fig:SM_SampleDrift}). Typically, the spread of the sampled points spans two vertical pixels, where the vertical length of each pixel is roughly equal to a diffraction-limited radius of $ \sim 500~\mathrm{nm}$ (Figure \ref{fig:SM_SpatialAnalysis}). Figure~\ref{fig:SM_SpatialAnalysis} shows the raw signal of the coherence decay $C(\tau, T)$ resolved per pixel at each temperature in the fitting range at criticality. The left panel at each temperature shows $C(\tau)$ for each individually tracked pixel with measurement error bars; the right panel shows the corresponding stray magnetic field image with the pixel positions indicated by color-matched labels. While some small pixel-to-pixel variation is visible, the per-pixel coherence curves are in general consistent within the measurement error bars, indicating that real spatial variation in the decoherence dynamics is weak on this length scale and does not meaningfully affect the critical exponents extracted from the pixel-averaged data.

\clearpage

\section{Supplementary Text}

\subsection{Temperature Range of Spin Echo Normalization}
\label{sec:SM_TemperatureNormalization}

At temperatures near the CrSBr phase transition, we observed a decrease in the pulsed ODMR contrast in the interior region of the CrSBr flake relative to the contrast outside the flake (Figure \ref{fig:SM_ContrastImages}a). This effect is consistent with the formation of sub-diffraction-limited static magnetic domains   in the flake interior near $T_C$ as reported in few-layer CrSBr by Tschudin \textit{et al.} with scanning NV magnetometry~\cite{tschudinImaging2024}, which we would expect to broaden the inhomogeneous linewidth ($\Gamma$) of the NV ensemble probed by the diffraction-limited laser spot in our experiment. Changes in the ratio of the Rabi frequency $\Omega$ to the inhomogeneous linewidth $\Gamma$ with temperature could affect the signal contrast in spin echo experiments \cite{dreauAvoiding2011}, complicating the comparison of data across a broad temperature range.

In the following analysis we show that, based on the variation in pulsed ODMR contrast in the temperature range of $\sim130-140$~K, we do not expect spin echo contrast to vary significantly enough to affect the analysis of the decoherence dynamics discussed in the main text. In particular, we show that the measured pulsed ODMR contrast (Figure \ref{fig:SM_ContrastImages}a.) can be used as a proxy to determine the ratio $\Omega/\Gamma$ at all temperatures. We then perform numerics to simulate the spin echo protocol at each value of $\Omega/\Gamma$ and demonstrate that the maximum effect on the normalized spin echo contrast is negligible. 

We begin by comparing the pulsed ODMR contrast in the interior and exterior regions of the flake as a function of temperature. Since the Rabi frequency $\Omega$ is fixed across all measurements, this comparison will allow us to evaluate temperature-dependent changes in the inhomogeneous broadening $\Gamma$ within the flake relative to the (temperature-independent) reference value outside the flake $\Gamma_0$.  We use our drift correction method (Section \ref{sec:SM_sampleDrift}) to track the spread of spatial points sampled over the total course of spin echo measurements at $137.5\:$K, which is shown by the red dots overlaid on the pulsed ODMR contrast image in Figure \ref{fig:SM_ContrastImages}a. From the spread of points we  determine the probability that a pixel in the image was sampled during the spin echo measurement (Figure \ref{fig:SM_ContrastImages}b) and calculate an average pulsed ODMR contrast for those points, obtained at $\Omega/\Gamma$. We also calculate an average pulsed ODMR contrast for a region of pixels off of the flake obtained at $\Omega/\Gamma_0$ (Figure \ref{fig:SM_ContrastImages}c) and determine the ratio of on-flake/off-flake pulsed ODMR contrast. This process is repeated for the data taken at all temperatures and the ratio of on-flake/off-flake pulsed ODMR contrast as a function of temperature is shown in Figure \ref{fig:SM_ContrastImages}d. 

Next, we demonstrate that the ODMR contrast ratio plotted in Figure \ref{fig:SM_ContrastImages}d is a valid proxy for the ratio $\Gamma/\Gamma_0$. We begin by simulating the Rabi contrast for an inhomogeneously broadened NV ensemble. We model the two-level dynamics of the pseudo-spin-$1/2$ subspace of the NV ground state \cite{dohertyNitrogenvacancy2013, schweigerPrinciples2001} using \textsc{Qiskit Dynamics}~\cite{puzzuoliQiskit2023}, where we define a rotating frame Hamiltonian in the $S_z$ basis with eigenstates $|{\alpha}\rangle=\begin{pmatrix}1\\0\end{pmatrix}$ and $|{\beta}\rangle=\begin{pmatrix}0\\1\end{pmatrix}$:
\begin{equation}
    H = h_z s_z + h_x s_x
\end{equation}
where $s_z$ and $s_x$ are the Pauli spin-$1/2$ operators, $h_x=2\pi \Omega$  (where $\Omega$ is the Rabi frequency),  and $h_z$  is the off-resonance Zeeman energy. In each simulation the initial state is set to $|{\psi_0}\rangle$=$|{\alpha}\rangle$ and the state is evolved over time (in units of $t_{2\pi}=1/\Omega$).  To simulate Rabi oscillations for different ratios of $\Omega/\Gamma$, we set the Rabi frequency $\Omega=1$ for all simulations and randomly select values of $h_z$ from a normal distribution of  width $2\pi\Gamma$. For a given normal distribution, we simulate the dynamics for $>10^3$ values of  $h_z$, calculating the expectation value $\langle S_z \rangle= \frac{1}{2}(|c_\alpha|^2 - |c_\beta|^2)$ at each time point for each value of $h_z$. We then average the results over all values of $h_z$ to compute the ensemble averaged expectation value $\langle S_z \rangle_{\text{ens}}$,  and take $\frac{1}{2}+\langle S_z \rangle_{\text{ens}}$ as the simulated contrast \cite{barrySensitivity2020}. Simulations are performed for a range of $\Gamma$ from $0.1-10$. Figure \ref{fig:SM_TempDepOmegaGammaRatio}a shows the simulated Rabi oscillations for two values of $\Omega/\Gamma$, which are fit to a function of the form:
\begin{equation}
Ae^{-t/T_\text{Rabi}}\cos(2\pi\Omega t) + B
\label{eq:RabiOsc}
\end{equation}
where $T_\text{Rabi}$ is a time constant for the Rabi decay. Figure \ref{fig:SM_TempDepOmegaGammaRatio}b shows $T_\text{Rabi}$ (in units of $t_{2\pi}$) determined for a wide range of $\Omega/\Gamma$. 
Figures \ref{fig:SM_TempDepOmegaGammaRatio}c,d show the Rabi oscillations measured at on-flake and off-flake positions  at $137.5$~K. By fitting the data in Figures \ref{fig:SM_TempDepOmegaGammaRatio}c,d we extract a decay time $T_\text{Rabi}$  in units of $t_{2\pi}$, from which we can determine values of  $\Omega/\Gamma$ by interpolating Figure \ref{fig:SM_TempDepOmegaGammaRatio}b. We determine a value of $\Omega/\Gamma_0=3.1$ for the off-flake position and $\Omega/\Gamma=1.03$ for the on-flake position at $137.5$~K. 
Next we determine the contrast for a $\pi$-pulse by evaluating $\frac{1}{2}+\langle S_z \rangle_{\text{ens}}$  at $t=0.5$ from the Rabi oscillation simulations. Figure \ref{fig:SM_TempDepOmegaGammaRatio}e shows the simulated $\pi-$pulse contrast as a function of  $\Omega/\Gamma$. Using  the on-flake $\Omega/\Gamma$ and off-flake  $\Omega/\Gamma_0$ values at $137.5 K$ determined above, we use Figure \ref{fig:SM_TempDepOmegaGammaRatio}e to estimate the value for the ratio of on-flake/off-flake $\pi-$pulse ODMR contrast. We determine an estimated value of $0.58$, which is in good agreement with the experimental value of $0.64$  at $137.5$~K  (Figure \ref{fig:SM_ContrastImages}d).
Assuming that the off-flake $\Omega/\Gamma_0=3.1$ is temperature independent, we then use the ratios of measured on-flake/off-flake pulsed ODMR contrast from Figure \ref{fig:SM_ContrastImages}d to determine the values of the on-flake  $\Omega/\Gamma$  ratio at all temperatures (Figure \ref{fig:SM_TempDepOmegaGammaRatio}f). 
In order to understand the effect of the trend in Figure \ref{fig:SM_ContrastImages}d on spin echo measurements, we determine an approximate relationship between the on-flake/off-flake pulsed ODMR contrast and initial spin contrast by performing a master equation simulation of spin echo as a function of $\Omega/\Gamma$. We define a time-dependent Hamiltonian to simulate coherent driving in a spin echo $\frac{\pi}{2}_{(x)}\rightarrow \pi_{(y)} \rightarrow \frac{\pi}{2}_{(x, -x)}$ pulse sequence with zero inter-pulse delay:
\begin{equation}
H = h_z s_z + H_1(t)
\end{equation}
\begin{displaymath}
\begin{split}
H_1(t) =2\pi\Omega \begin{cases} 
      s_x & 0\leq t \leq \frac{1}{4} \\
      s_y & \frac{1}{4}\leq t\leq \frac{3}{4} \\
      s_{x,-x} & \frac{3}{4}\leq t\leq 1 
   \end{cases}
\end{split}
\end{displaymath}
Here two simulations are performed: one with a final $s_x$ rotation and one with a final $s_{-x}$ rotation. Using a similar method to model inhomogeneous broadening as described above, we fix $\Omega=1$ and perform simulations over randomly selected values of  $h_z$  taken from a normal distribution of width $2\pi\Gamma$, and evaluate the results for different values of  $\Gamma$. In each simulation the initial spin echo contrast is defined as $\langle S_z \rangle_{echo}=\langle S_z \rangle_x-\langle S_z \rangle_{-x}$, evaluated at $t=1$ (in units of $t_{2\pi}$).  For each value of $\Gamma$  we average the contrast for all values of $h_z$ to determine the ensemble averaged spin echo contrast  $\langle S_z \rangle_{{\text{ens}}_{echo}}=\langle S_z \rangle_{ens_{x}}-\langle S_z \rangle_{ens_{-x}}$. Figure \ref{fig:SM_SpinEchoContrast}a shows the simulated initial spin echo contrast as a function of $\Omega/\Gamma$. Using the temperature-dependent values of the on-flake $\Omega/\Gamma$ determined above (Figure \ref{fig:SM_TempDepOmegaGammaRatio}f), we interpolate Figure \ref{fig:SM_SpinEchoContrast}a to estimate the fraction of spin echo contrast lost at each temperature in our experiment (Figure \ref{fig:SM_SpinEchoContrast}b).

Finally, we verify that the effect of the $\Omega/\Gamma$ ratio on spin manipulation does not affect the functional form of the observed decoherence dynamics by simulating a   $\frac{\pi}{2}_{(x)}\rightarrow \tau\rightarrow \pi_{(y)} \rightarrow \tau \rightarrow \frac{\pi}{2}_{(x, -x)}$ spin echo sequence with pure dephasing and variable inter-pulse delay $\tau$. Here we define a Hamiltonian:

\begin{equation}
H = h_z s_z + H_1(t)
\end{equation}
\begin{displaymath}
\begin{split}
H_1(t) =2\pi\Omega \begin{cases} 
      s_x & 0\leq t \leq \frac{1}{4} \\
      s_y & \tau + \frac{1}{4}\leq t\leq \tau + \frac{3}{4} \\
      s_{x,-x} & 2\tau+\frac{3}{4}\leq t\leq2\tau+1 
   \end{cases}
\end{split}
\end{displaymath}
with $t$ and $\tau$ in units of $t_{2\pi}$. To simulate pure dephasing we define a relaxation operator $\sqrt{\Gamma_2}s_z$ for the dissipative term of the Lindblad equation which causes the off-diagonal elements of the density matrix to decay  at a rate $2\Gamma_2$. Simulations for a range of  $\Omega/\Gamma$  are performed in the same way as above for the initial spin echo contrast, with $\Omega=1$  and  $\Gamma_2=0.3$ fixed for all simulations. Figure \ref{fig:SM_SpinEchoContrast}c shows the results of simulations for three different values of $\Omega/\Gamma$. Figure \ref{fig:SM_SpinEchoContrast}d shows the simulations normalized to the value of $\langle S_z \rangle_{echo}$ at $\tau=0$, which shows that the measured decoherence dynamics in the $\frac{\pi}{2}_{(x)}\rightarrow \tau\rightarrow \pi_{(y)} \rightarrow \tau \rightarrow \frac{\pi}{2}_{(x, -x)}$ spin echo pulse sequence are not changed by the effect of  $\Omega/\Gamma$ on spin manipulation. The contrast correction to the observed decoherence dynamics therefore reduces to a single temperature-dependent scalar $f(T)$:
\begin{equation}
C^\text{corr}(\tau, T) = f(T)\cdot C_\text{raw}(\tau, T)
\end{equation}
Figure \ref{fig:SM_SpinEchoContrast}b shows that this scalar varies by $\lesssim 5\%$ across the full temperature range of 130–142 K. Given that this variation is negligible compared to the order-of-magnitude changes in decoherence rate driven by critical fluctuations near $T_C$, we do not expect the contrast correction to have a meaningful effect on the extracted critical exponents and overall conclusions of our analysis.

\subsection{Critical Dynamics Analysis}
\label{sec:SI_EnsembleCoherece}

\subsubsection{Simulation Methods and Definitions}
We simulated the ensemble decoherence by generalizing the results in Machado \textit{et al.} \cite{machadoQuantum2023} to carefully account for critical dynamics beyond mean-field theory. We start with the calculation of $\langle\phi^2\rangle_{\ell}$ for a single NV center at a depth $\ell$  from the 2D flake, given by:
\begin{equation} \label{eq:phi2}
\langle\phi^2\rangle_\ell =\int_{\omega_L}^{\omega_U} W_\tau(\omega)\,\frac{d\omega}{2\pi}\int_{0}^{1/a} W_\ell(q)S(q,\omega) \,\frac{dq}{2\pi} 
\end{equation}
with:
\begin{equation}
W_\tau(\omega) = \left(\frac{2g_e\mu_B}{\hbar}\right)^2\frac{\sin^4{(\frac{\omega\tau}{4})}}{\omega^2} 
\end{equation}

where $g_e$ is the electron g-factor and $\mu_B$ is the Bohr magneton, and:
\begin{equation}
W_\ell(q) = \frac{(\mu_0\mu_Bg_eS)^2}{4a^4}q^3e^{-2q\ell}
\end{equation}
where $\mu_0$ is the vacuum permeability, $S=\frac{3}{2}$ is the electron spin for Cr$^{3+}$\cite{ziebelCrSBr2024}, and $a=4.7485\: \mathrm{\mathring{A}}$ is taken as the CrSBr lattice constant \cite{lopez-pazDynamic2022}.

The decoherence of the whole ensemble is obtained by averaging the coherence signal of spins at different depths. The resulting ensemble signal is given by:
\begin{equation}
    C_{ens} = \int d\ell ~\rho(\ell)e^{-\langle \phi^2\rangle_\ell/2}    
\end{equation}
where $\rho(\ell)$ is the density of NVs as a function of implantation depth $\ell$, and $-\langle \phi^2\rangle_\ell$ is the phase variance of a single NV at depth $\ell$, Eq.~\ref{eq:phi2}.

The low temperature correlation length, which is on the scale of the lattice, is defined as $\xi_0=2a$ \cite{machadoQuantum2023}. We take $J = 3.36 ~\text{meV}$, which is determined based on the sum of contributions coming from the nearest and next nearest exchange interactions along the $\hat{b}$ axis  \cite{scheieSpin2022}. 

We note that within the parameters considered in this work, the particular choice of $J$ enters with $\omega_0$; that is, the two parameters are correlated and primarily determine the overall strength of the noise. Specifically,  the scaling function in the small-frequency limit scales as $1/\omega_0 J$; therefore our method of allowing $\omega_0$ to vary as a free parameter in the fits allows us to model the correct strength of the noise irrespective of the exact value of $J$.
To check whether or not we are probing an appropriate small frequency limit, we note that $\omega_0\gtrsim 3~\mathrm{GHz}$, which sets the width of the structure factor, is much greater than the position of the frequency filter function, which is $\lesssim 10~\text{ MHz}$.

\subsubsection{General constraints on the structure factor $S({\bm q},\omega)$} 
The dynamical structure factor $S({\bm q},\omega)$ for spins is calculated using the \textit{classical} fluctuation-dissipation theorem that relates the structure factor to the dynamic susceptibility $\chi({\bm q},\omega)$, as the scaling theory for critical phenomena is typically framed in terms of the susceptibility.
\begin{align}
S({\bm q},\omega) &= \frac{2k_BT}{\omega} \text{Im}[\chi({\bm q},\omega)], \\
S({\bm q}) &= \int_{-\infty}^{\infty} \frac{d\omega}{2\pi} S({\bm q},\omega) = k_B T \chi({\bm q})
\label{eq:rel}
\end{align}
where $\chi({\bm q}) = \chi({\bm q},\omega = 0)$ is the static susceptibility.
Eq.~\eqref{eq:rel} holds because of Kramers-Kronig relations, which are exact and follow from causality.
\begin{equation}
\text{Re}[\chi({\bm q},\omega)] = \mathcal{P}\int_{-\infty}^{\infty} \frac{{d\omega^\prime}}{\pi} \frac{\text{Im}[\chi({\bm q},\omega^\prime)]}{\omega^\prime - \omega}, ~~~ \text{Im}[\chi({\bm q},\omega)] = -\mathcal{P}\int_{-\infty}^{\infty} \frac{{d\omega^\prime}}{\pi} \frac{\text{Re}[\chi({\bm q},\omega^\prime)]}{\omega^\prime - \omega}
\label{eq:KK}
\end{equation}
where $\mathcal{P}$ denotes the Cauchy principal value of the integral.
Specifically, focusing on $\omega = 0$ in the LHS of Eq.~\eqref{eq:KK}, we find that 
\begin{equation}
 \text{Re}[\chi({\bm q},\omega = 0)] = \chi({\bm q}) = \mathcal{P}\int_{-\infty}^{\infty} \frac{{d\omega^\prime}}{\pi} \frac{\text{Im}[\chi({\bm q},\omega^\prime)]}{\omega^\prime}
 \label{eq:KK2}
\end{equation}
It also follows from the quantum version of linear response theory (and continues to hold in the classical limit) that
\begin{equation}
\text{Re}[\chi({\bm q},\omega)] = \text{Re}[\chi({\bm q},-\omega)], ~~~ \text{Im}[\chi({\bm q},\omega)] = -\text{Im}[\chi({\bm q},-\omega)],
\label{eq:reim}
\end{equation}
which is consistent with Eq.~\eqref{eq:KK}.
Eq.~\eqref{eq:reim} in turn implies that $S({\bm q},\omega) = S({\bm q},-\omega)$ as $\text{Im}[\chi({\bm q},\omega)]/\omega$ will be an even function of $\omega$ (in the classical limit $\hbar \omega \ll k_B T$).

\subsubsection{Scaling form for the dynamical susceptibility}
While the previous constraints on $S({\bm q},\omega)$ and $\chi({\bm q},\omega)$ are quite general, we now specify their form when the system is near criticality.
In this regime, the scaling form for the (scalar) order parameter susceptibility takes the form:
\begin{equation}
\chi({\bm q},\omega) = q^{-2 + \eta} \chi_{\rm sc}\left(q\xi, \frac{\omega}{q^z} \right),
\end{equation}
where $\xi$ is the order parameter correlation length that diverges near the transition as $\xi  = \xi_0 |(T - T_c)/T_c|^{-\nu}$~[we address the BKT transition case below],  $\eta,\nu$ and $z$ are critical exponents, and $\chi_{sc}$ is a dimensionless scaling function.
The simplest extension of the scaling form, which follows from a time-dependent Ginzburg-Landau approach and preserves the Kramers-Kronig relations [see Eq.~\eqref{eq:reim}], is given by
\begin{equation}
\chi^{-1}({\bm q},\omega) = \chi^{-1}({\bm q}) - \frac{i \omega}{\Gamma({\bm q})},
\end{equation}
where $\Gamma({\bm q})$, a kinetic coefficient, defines a characteristic dissipative frequency scale $\omega_c({\bm q}) = \Gamma({\bm q})/\chi({\bm q})$ that determines how fast a spatial modulation in the order parameter at momentum ${\bm q}$ decays to the thermal state.
Crucially, $\Gamma({\bm q})$ should not be treated as a number but rather as a flowing coupling~\cite{hohenbergTheory1977}.
In the renormalization group setting, the energy and momenta scale as $q \to b q, ~ \omega \to b^z \omega$ for $b > 1$. 
Hence, if we demand scale invariance of the dynamical susceptibility $\chi({\bm q},\omega)$ at the critical point ($\xi \to \infty$), we fix the scaling of the kinetic coefficient as $1/\Gamma \to b^{2 - \eta - z}/\Gamma$. 
This leads us to the scaling form for the kinetic coefficient:
\begin{equation}
\Gamma({\bm q}) =(qa)^{-2 + \eta + z} \bar{\Gamma}(q\xi).
\label{eq:Gamma}
\end{equation}
where $\bar{\Gamma}(q\xi)$ is a dimensionless scaling function, and the power of $q$ used in Eq.~\eqref{eq:Gamma} is fixed by the desired scaling of $\Gamma({\bm q},\omega)$. 
In mean-field theory, $\eta = 0$ and $z = 2$, so $\bar{\Gamma}(q\xi) = \Gamma_0$ is fixed, i.e., it is the constant function.
However, generically we expect the critical behavior to be described by an interacting fixed point, better described in terms of the renormalized $\Gamma({\bm q})$ in Eq.~\eqref{eq:Gamma}. 

In order to fit with the experimental data, we need a functional form for $\bar{\Gamma}(q\xi)$: this may be anticipated from physical arguments. 
As one approaches the critical point, $\bar{\Gamma}(\infty)$ should approach a non-zero constant  $\Gamma_0$ independent of ${
\bm q}$ if the scaling hypothesis were to remain valid.
This is consistent with the intuition that at length-scales $q^{-1} \lesssim \xi$, the system already appears critical. 
On the other hand, at long-wavelengths $q^{-1} \gtrsim \xi$,  $\Gamma({\bm q})$ should approach a constant independent of $q$, so $\bar{\Gamma}$ must scale as $(q\xi)^{2 - \eta - z}$ for $q\xi \lesssim 1$.
One can interpolate between these limits by postulating a functional form for $\Gamma_0({\bm q},\xi)$:
\begin{equation}
\Gamma({\bm q}) = (qa)^{-2+\eta + z} \bar{\Gamma}(q\xi), \text{ where } \bar{\Gamma}(q\xi) = \Gamma_0 \tanh\left[(q\xi)^{2- \eta - z}\right]
\label{eq:GammaSc}
\end{equation}
This, in turn, leads to analytical scaling forms for the dynamical structure factor, which we turn to next. 

\subsubsection{2D Ising universality class}
The most generic static structure factor for a transition in the 2D Ising universality class can be written as
\begin{equation}
\chi^{-1}({\bm q}) = (J/a^2)^{\eta/2} [J({\bm q}^2 + \xi^{-2})]^{1-\eta/2},
\end{equation}
where we have introduced $(J/a^2)^{\eta/2}$ for dimensional consistency. 
Accordingly, we have
\begin{align}\label{eq:SM_StructureFactorSSB}
 S({\bm q},\omega) &= \frac{2 k_B T}{\omega} \, \text{Im}[\chi({\bm q},\omega)] \notag \\
 &=  \frac{2 k_B T (qa)^{-2+\eta+z} \bar{\Gamma}(q\xi)}{(J/a^2)^{\eta}[J ({\bm q}^2 + \xi^{-2})]^{2-\eta} [(qa)^{-2+\eta + z} \bar{\Gamma}(q\xi)]^2 + \omega^2},\\[10pt]
 & \text{where}~\xi = \xi_0 \bigg| \frac{T - T_c}{T_c} \bigg|^{-\nu} \notag
\end{align}
We note that this form of the structure factor is valid on both sides of the transition.

This functional form was used to compute the decoherence rate of a single NV in Fig.~\ref{fig:fig3}b, using the following parameters:
$z=1.68$, $\eta=0$, $\nu = 0.58$, NV depth $\ell=50~\mathrm{nm}$, and $\omega_0=3~\mathrm{GHz}$. The value of $J = 3.36 ~\text{meV}$ and $a=4.7485~\AA$ are set by the material.

\subsubsection{2D XY universality class}

One hypothesis for the critical phenomena in CrSBr is a 2D XY critical point (i.e. the BKT universality class), which is governed by the dynamics of topological defects called vortices. 
As a consequence, the BKT universality class has crucial differences from the 2D Ising universality class --- the correlations of the XY order parameter have drastically different functional forms on the two sides of the transition. 

In this subsection, we describe the structure factor $S_{\text{BKT}}(\bm{q}, \omega)$ of this critical point in a way that can be utilized in Eq.~\ref{eq:phi2} to compute the decoherence rate of the spin ensemble.

For $T \leq T_{\rm BKT}$, the system transitions to a quasi-long-range ordered (QLRO) phase with power law correlations $r^{-\eta(T)}$, with an exponent $\eta(T)$ that explicitly depends on temperature.
In the scaling regime close to the critical point, $\eta(T)$ can be written in terms of the renormalized spin stiffness $\rho_s(T) \approx \rho_s(T_{\rm BKT}^{-})(1 + b \sqrt{1 - T/T_{\rm BKT}})$ as~\cite{chaikinPrinciples1995}
\begin{align}
    \eta(T) &= \frac{k_B T}{2\pi \rho_s(T)} = \underbrace{\left[  \frac{k_B T_{\rm BKT}}{2\pi \rho_s(T_{\rm BKT}^{-})} \right]}_{=\frac{1}{4}} \frac{T}{T_{\rm BKT}(1 + b \sqrt{1 - T/T_{\rm BKT}})} \\
    &= \frac{T}{4T_{\rm BKT}} \frac{1}{(1 + b \sqrt{1 - T/T_{\rm BKT}})}
\end{align}
As the temperature is reduced, the correlations decay more slowly and $\eta(T)$ becomes smaller, with $\eta(T \to 0^+) \to \frac{k_B T}{2\pi J}$. 
In this regime, one cannot meaningfully define a correlation length, and the entire phase is critical (in the sense that it has power law correlations). 
Hence, we may write
\begin{equation}
\chi^{-1}({\bm q},\omega) = \chi^{-1}({\bm q}) - \frac{i \omega}{\Gamma({\bm q})} = (J/a^2)^{\eta(T)/2} [J {\bm q}^2 ]^{1-\eta(T)/2} - \frac{i\omega}{\Gamma({\bm q})}
\end{equation}
where $\Gamma({\bm q})$ is a function of the momentum (but not of the correlation length $\xi$ since it is not well-defined for $T \leq T_{\rm BKT}$). 
Therefore, the scaling form for $\Gamma({\bm q})$ is given by
\begin{equation}
\Gamma({\bm q}) = (qa)^{-2+ \eta + z}\Gamma_0
\end{equation}
Accordingly, the structure factor is given by
\begin{align}
S({\bm q},\omega) &= \frac{2 k_B T \Gamma({\bm q})}{(J/a^2)^{\eta(T)} (J {\bm q}^2)^{2-\eta(T)} [\Gamma({\bm q})]^2 + \omega^2} \\
&= \frac{2k_B T(qa)^{-2+z+\eta(T)} \Gamma_0}{(J/a^2)^{\eta(T)} (J {\bm q}^2)^{2-\eta(T)} [(qa)^{-2+z+\eta(T)} \Gamma_0]^2 + \omega^2}
\end{align}

For the thermally disordered phase ($T > T_{\rm BKT}$), the correlations decay exponentially as $e^{-r/\xi(T)}$ at large distances. Additionally, for $T > T_{\rm BKT}$, the power-law divergence of the correlation length $\xi(T)$ as a function of $T - T_c$ is replaced by an exponential dependence for the BKT transition, leading to the following expression for the structure factor for $T > T_{\mathrm{BKT}}$:
\begin{align}
S({\bm q},\omega) &= \frac{2 k_B T}{\omega} \, \text{Im}[\chi({\bm q},\omega)] \notag \\
&=  \frac{2 k_B T (qa)^{-2+\eta+z} \bar{\Gamma}(q\xi)}{(J/a^2)^{\eta}[J ({\bm q}^2 + \xi^{-2})]^{2-\eta} [(qa)^{-2+\eta + z} \bar{\Gamma}(q\xi)]^2 + \omega^2}\\[10pt]
&\text{where} \quad \xi(T) = \xi_0 \, e^{b \sqrt{\frac{T_{\mathrm{BKT}}}{T - T_{\mathrm{BKT}}}}} \notag
\end{align}

For $T > T_{\rm BKT}$, we utilize a phenomenological form for $\eta(T)$ that smoothly interpolates between its critical value of $1/4$ at $T = T_{\rm BKT}$ and an infinite-temperature value $1/2$ where correlations are purely exponentially decaying. We utilize the following extrapolation:
\begin{equation}
    \eta_{\mathrm{BKT}}(T) = \frac{1}{4} \left( 1+ \sqrt{T/T_{\mathrm{BKT}} - 1} \right) e^{-(T-T_{\mathrm{BKT}})/T_0}   + \frac{1}{2} \left(1-e^{-(T-T_{\mathrm{BKT}})/T_0}\right),
\end{equation}
with $T_0=30~\mathrm{K}$ corresponding to the temperature scale away from criticality where the system is paramagnetic. This temperature scale is set to be much larger than the critical window explored in this work, i.e., $T_0 \gg T-T_{\text{BKT}}$. 

This functional form was used to compute the decoherence rate of a single NV in Fig.~\ref{fig:fig3}b, using the following parameters:
$b=0.5$, $z = 2$, NV depth $\ell=50~\mathrm{nm}$, and $\omega_0=5~\mathrm{GHz}$. The value of $J = 3.36 ~\text{meV}$ and $a=4.7485~$\AA ~ are set by the material.

\subsubsection{Choice of fitting parameters and robustness of our observations}
\label{sec:SM_parameterChoice}

Throughout our work, we have made informed decisions for some of the parameters considered in our fitting procedure.
This is an attempt to simplify our fitting procedure and thus provide more accurate and transparent measurements of the critical exponents. In this section we  highlight those decisions and discuss the robustness of our results (namely that the observed critical fluctuations are distinct from those arising from the conventional short-range Ising criticality) to our particular choices.
The elements we consider are:
\begin{itemize}
    \item The normalization temperature $T_{\text{Norm}}$: 
    to remove the decoherence dynamics arising from external sources to the material sample, we have chosen a reference temperature $T_{\text{Norm}}=130~\mathrm{K}$ whose decoherence dynamics we use to normalize the data before fitting to the decoherence dynamics simulation data.
    We perform the same analysis using a different temperature $T_{\text{Norm}}=132~\mathrm{K}$, away from the critical regime.
    The best fits, and reduced $\chi^2$ landscapes are summarized in Fig.~\ref{fig:SM_robustness} 1st row, in qualitative agreement with the work presented in the main text.
    \item The temperature range:
    when fitting to multiple temperatures near the critical point $T_C=139~\mathrm{K}$, there is a choice of the range of temperatures whose datasets we consider. 
    In the main text, we have focused on $T \in [137, 138.25]~\mathrm{K}$.
    This was motivated by two constraints: 
    the validity of our analysis is contingent on the structure factor following a universal behavior parameterized by the critical exponents (which occurs close to the phase transition), and, close to criticality, the decoherence dynamics are challenging to accurately measure owing to its fast decay.
    For a higher temperature range, $138.5~\mathrm{K}$ in Fig.~\ref{fig:SM_robustness} 2nd row, the decoherence dynamics overlap across different temperatures.
    For a lower temperature range, $136.75~\mathrm{K}$ in Fig.~\ref{fig:SM_robustness} 3rd row, the decoherence dynamics start to strongly deviate from the universal scaling prediction.
    Regardless of these considerations, the resulting poorer fits still demonstrate the qualitative features emphasized in the main text.\\
    \item The critical temperature $T_C$:
    Our choice of critical temperature is based on the peak in the decoherence rate.
    However, the two greatest decoherence rates are consistent with one another, implying some important uncertainty in the position of the critical temperature that was not taken into account in the main text analysis.
    In Fig.~\ref{fig:SM_robustness} 4th row, we consider our fitting pipeline assuming the critical temperature is $138.75~\mathrm{K}$, observing an equally good fit to that presented in the main text, and whose conclusion matches our observations.
    \item The correlation function critical exponent $\eta$:
    In our analysis, we implicitly assumed that the correlation function critical exponent $\eta=0$, despite it being $1/4$ for the two-dimensional Ising point. 
    This simplification was motivated by the small value that $\eta$ usually takes in higher dimensional, non-solvable, critical points.
    In Fig.~\ref{fig:SM_robustness} 5th row, we test this assumption by considering the more general case for the structure factor with $\eta=1/4$ (see Eq.~\ref{eq:SM_StructureFactorSSB}). 
    Crucially, we observe that such large $\eta$ strongly modifies the early time decoherence dynamics, and leads to decoherence dynamics that are {\bf not} well captured by our model, suggesting that our assumption of small $\eta$ is well-motivated.    
\end{itemize}

\clearpage
\newpage

\clearpage

\clearpage
\clearpage
\clearpage
\clearpage

\clearpage
\begin{figure}
\centering
\includegraphics[width=0.75\textwidth]{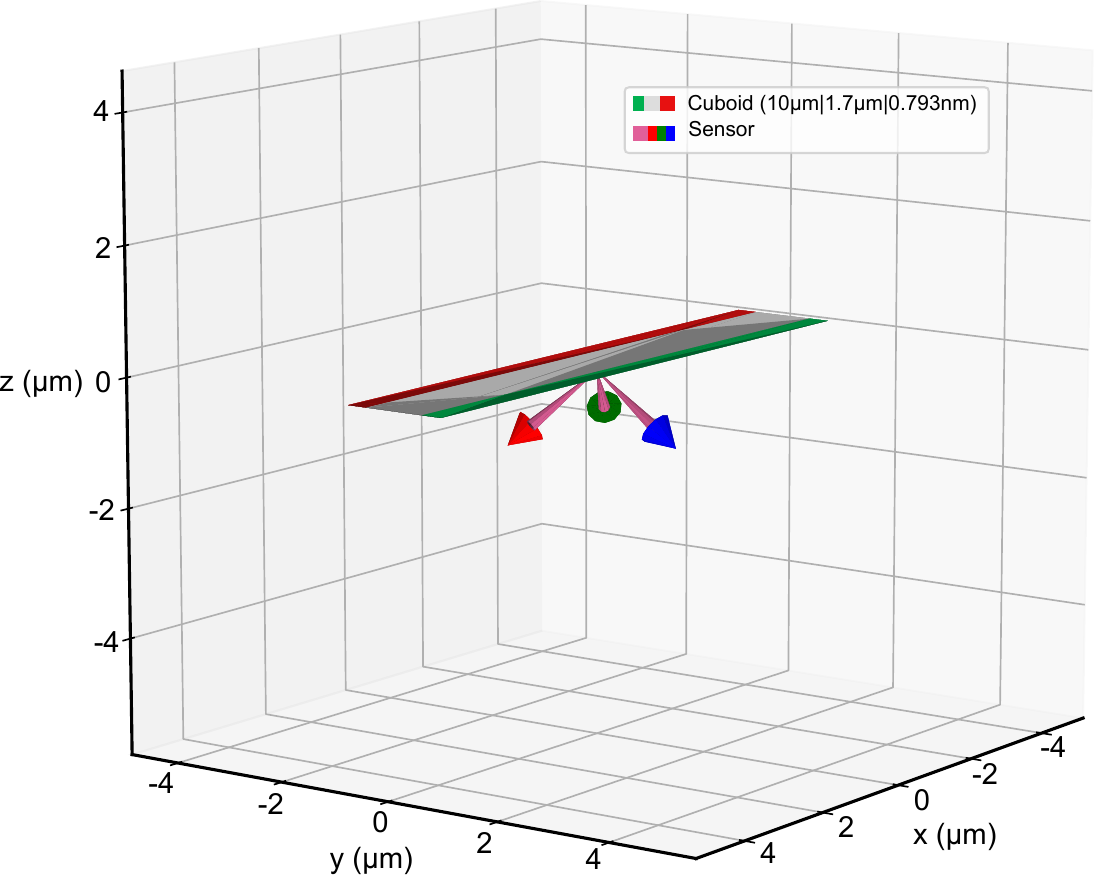}
\caption{\textbf{$B_{\text{stray}}$ simulation details.}~We used \textsc{Magpylib} \cite{ortnerMagpylib2020} to simulate the stray magnetic fields from one uncompensated ferromagnetic monolayer of CrSBr. The geometry of the simulation setup is shown above. We defined the CrSBr flake as a cuboid with $x$, $y$ dimensions of $10 \:\mathrm{\mu m} \times 1.7 \:\mathrm{\mu m}$ and a $z$ dimension of $0.793~$nm corresponding to the thickness of one monolayer of CrSBr \cite{lopez-pazDynamic2022}. The magnetization vector is oriented along the $y$ axis and its magnitude (the remanence) is calculated from the CrSBr unit cell magnetization at $100 \:$K reported by L\'{o}pez-Paz \textit{et al.}\cite{lopez-pazDynamic2022} with the CrSBr unit cell volume. Experimentally, the CrSBr flake was oriented roughly so that the rectangular edges of the flake were parallel with the square edges of the diamond substrate during the transfer process such that $x$ and $y$ axes of the flake  in the simulation are taken to point along $[110]$ directions of the diamond unit cell, while the $z$ axis corresponds to the $[100]$ direction. We  rotate the sensor coordinate system in the simulation such that the $z$-axis of the sensor (blue arrow in the figure above) in the new coordinate system points in a direction corresponding to the $[111]$ direction of the diamond unit cell, i.e. along the NV spin quantization axis. The sensor is  placed at a distance $\ell$  below the flake and scanned in the $x$-$y$ plane and the B-field projection onto the sensor $z$-axis is calculated to obtain a simulated image $B_{sim}(x,y,\ell)$. We then perform simulations of  $B_{sim}(x,y,\ell)$ over the range of $\ell_{NV}$ spanned by the NV depth probability distribution function $\rho(\ell_{NV})$ (Fig \ref{fig:SM_NVDensity}) and calculate an ensemble averaged image $B_{sim_{\text{ens}}}(x,y)=\int_{}^{}B_{sim}(x,y,\ell_{NV})\rho(\ell_{NV})\,d\ell_{NV}$. Finally, the ensemble averaged image $B_{sim_{\text{ens}}}(x,y)$ is convolved with a 2D Gaussian representing the optical diffraction-limited width for a beam waist radius of $\sim 500 \:$nm to generate the image shown in Fig \ref{fig:fig1}e of the main text.}
\label{fig:SM_MagpylibSim}
\end{figure}

\clearpage
\begin{figure}
\includegraphics[width=1\textwidth]{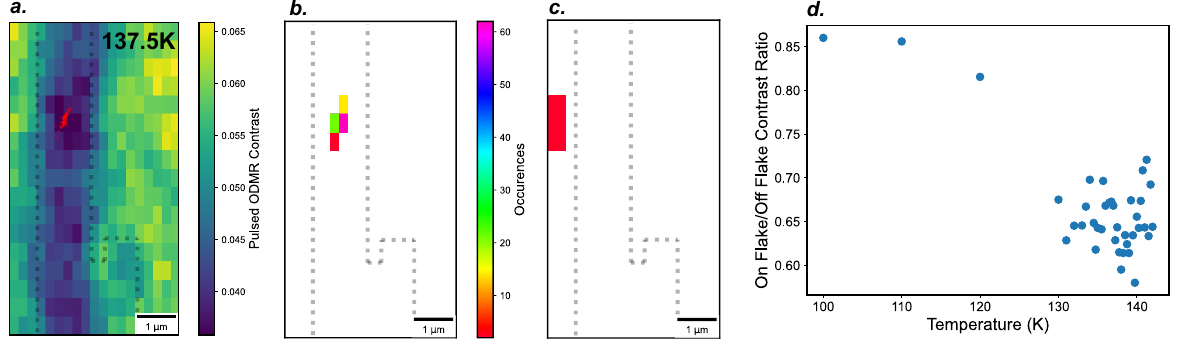}
\caption{\textbf{Temperature dependence of on-flake vs. off-flake pulsed ODMR contrast.}\\(a) Pulsed ODMR contrast image (see Section \ref{sec:SM_sampleDrift}) with points sampled during spin echo measurements overlaid in red. (b) Histogram of points sampled during spin echo measurements. (c) Pixels selected to evaluate ``off-flake" pulsed ODMR contrast. (d) Ratio of pulsed ODMR contrast from ``on-flake" vs. ``off-flake" positions as a function of temperature. }
\label{fig:SM_ContrastImages}

\end{figure}

\begin{figure}
\includegraphics[width=1\textwidth]{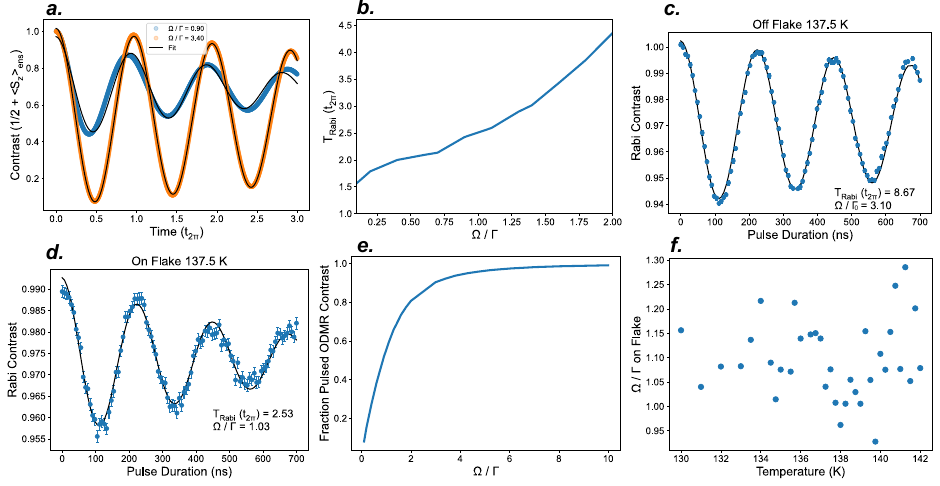}
\caption{\textbf{Determining the temperature dependence of $\Omega/\Gamma$ across the CrSBr phase transition.}~(a) Simulated Rabi contrast for two values of $\Omega/\Gamma$ with fits to Eq (\ref{eq:RabiOsc}). (b)  Simulated $T_\text{Rabi}$ as a function of $\Omega/\Gamma$. (c) \& (d) Rabi oscillations measured at positions on and off the CrSBr flake at $137.5 \:K$, from which $T_\text{Rabi}$, $\Omega/\Gamma$, and $\Omega/\Gamma_0$ are determined. (e) Simulated $\pi-$pulsed ODMR contrast as a function of $\Omega/\Gamma$. (f) $\Omega/\Gamma$ at the position on the CrSBr flake sampled in spin echo measurements as a function of temperature.}
\label{fig:SM_TempDepOmegaGammaRatio}
\end{figure}

\clearpage
\begin{figure}
\includegraphics[width=1\textwidth]{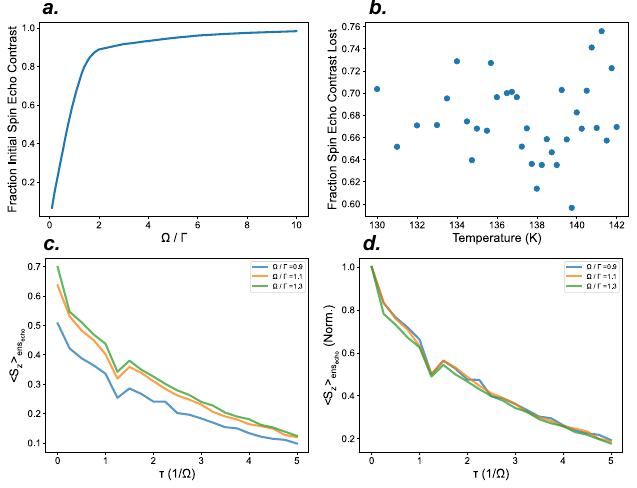}
\caption{\textbf{Determining temperature dependence of spin echo contrast across the CrSBr phase transition.}~(a) Simulated initial spin echo contrast as a function of $\Omega/\Gamma$. (b) Spin echo contrast lost as a function of temperature determined from the values of $\Omega/\Gamma$ at the position on the CrSBr flake sampled in spin echo measurements. (c) \& (d) Simulated spin echo coherence decays for different values of  $\Omega/\Gamma$.}
\label{fig:SM_SpinEchoContrast}

\end{figure}

\newpage
\begin{figure}
\centering
\includegraphics[width=0.65\textwidth]{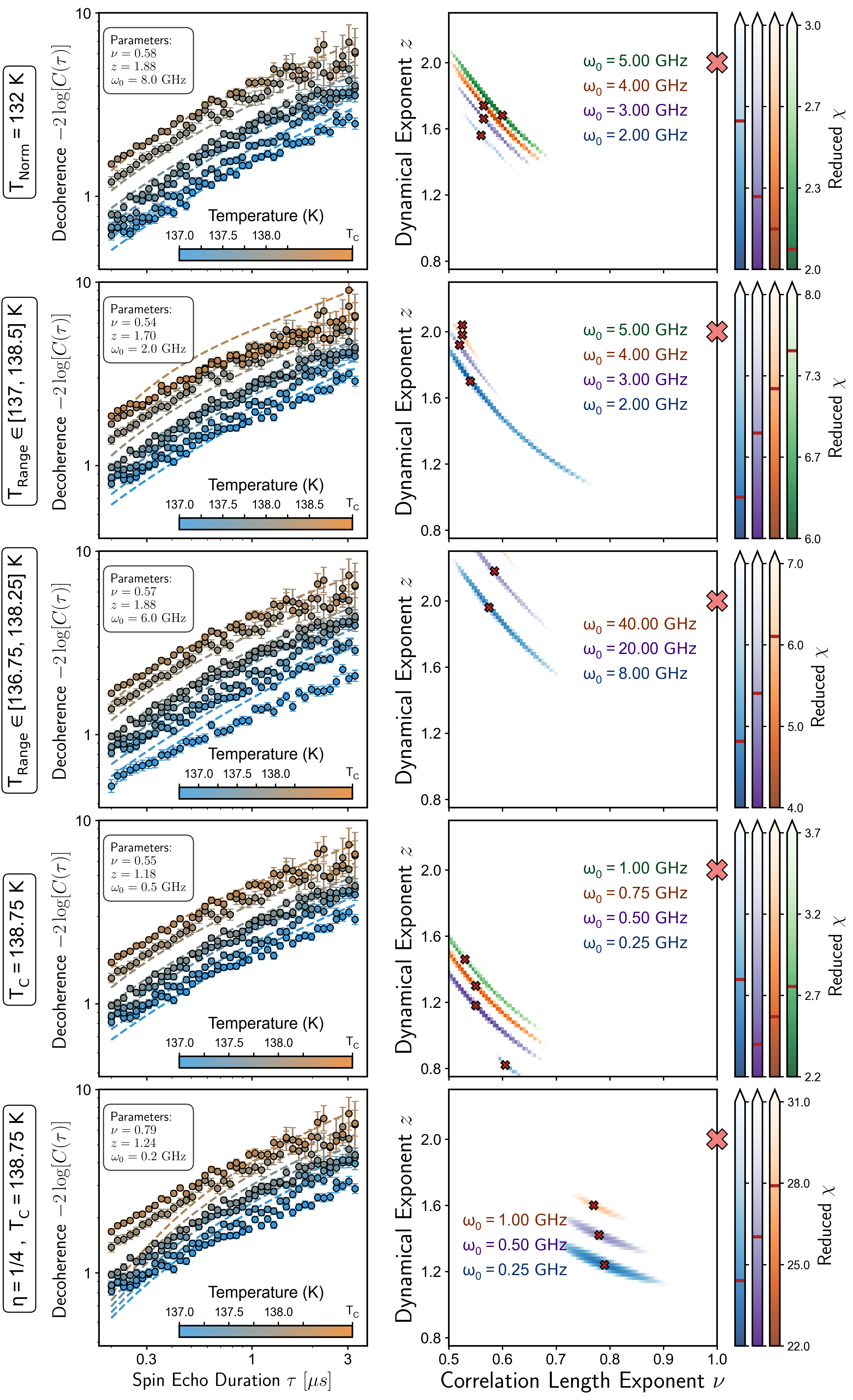}
\caption{Robustness of the critical exponent extraction across the different choices described in Sec.~\ref{sec:SM_parameterChoice}: normalization temperature, upper limit of critical regime, lower limit of critical regime, critical temperature, and choice of setting $\eta=0$ with critical temperature set to be $138.75$~K. }
\label{fig:SM_robustness}
\end{figure}

\newpage
\begin{figure}
\centering
\includegraphics[width=0.65\textwidth]{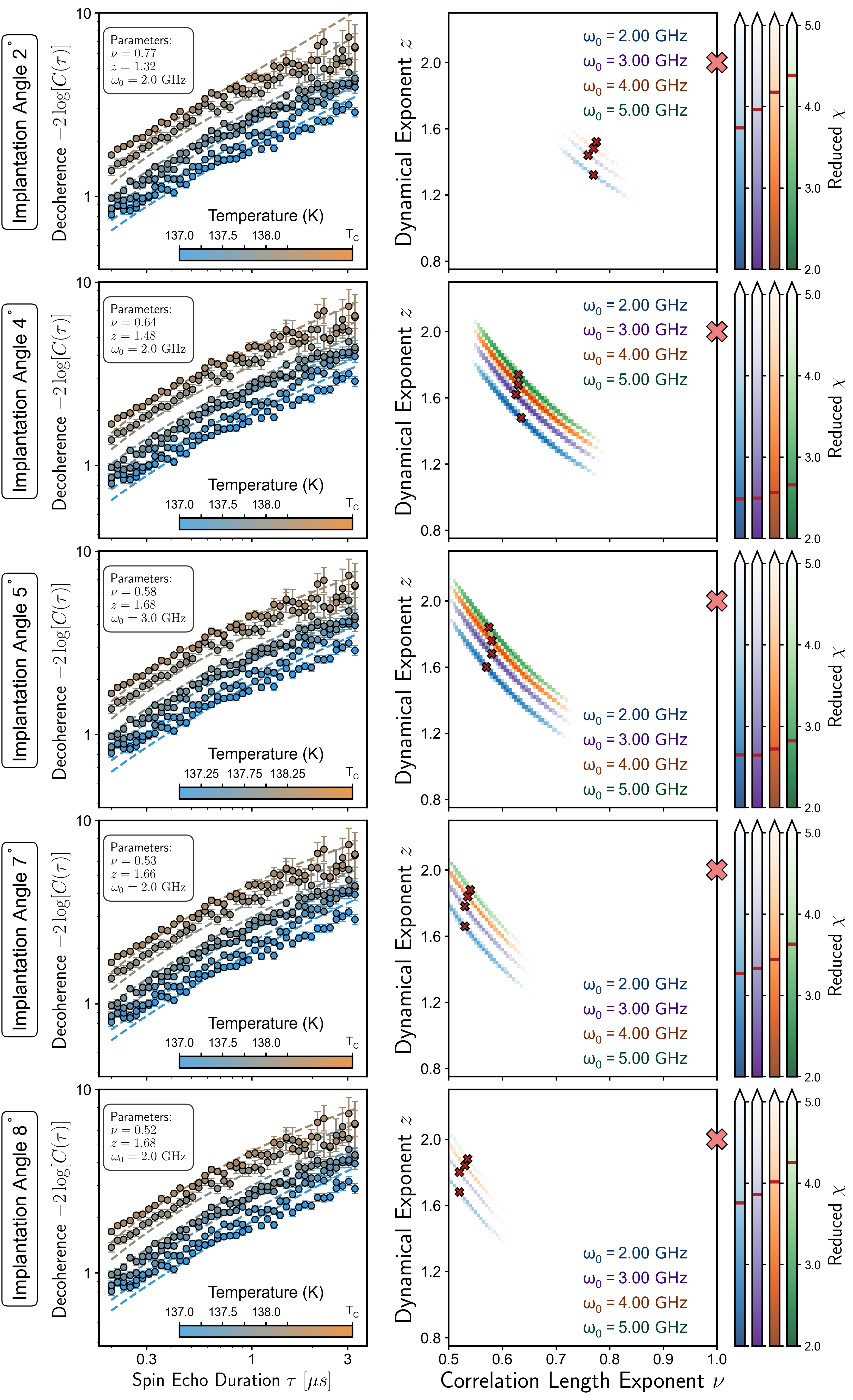}
\caption{Impact of implantation uncertainty on the extracted critical exponents $z$ and $\nu$. 
The uncertainty in the implantation angle $5^\circ \pm 3^\circ$, modifies the predicted NV depth distribution which affects the details of the decoherence profile. 
However, across the entire range, we observe that the extracted critical exponents remain well separated from those predicted by the short-range Ising universality class. 
Notably, the global best fits occur for implantation angle around the expected $5^\circ$.}
\label{fig:SM_ImplantationAngle}
\end{figure}

\newpage
\begin{figure}
\centering
\includegraphics[width=0.6\textwidth]{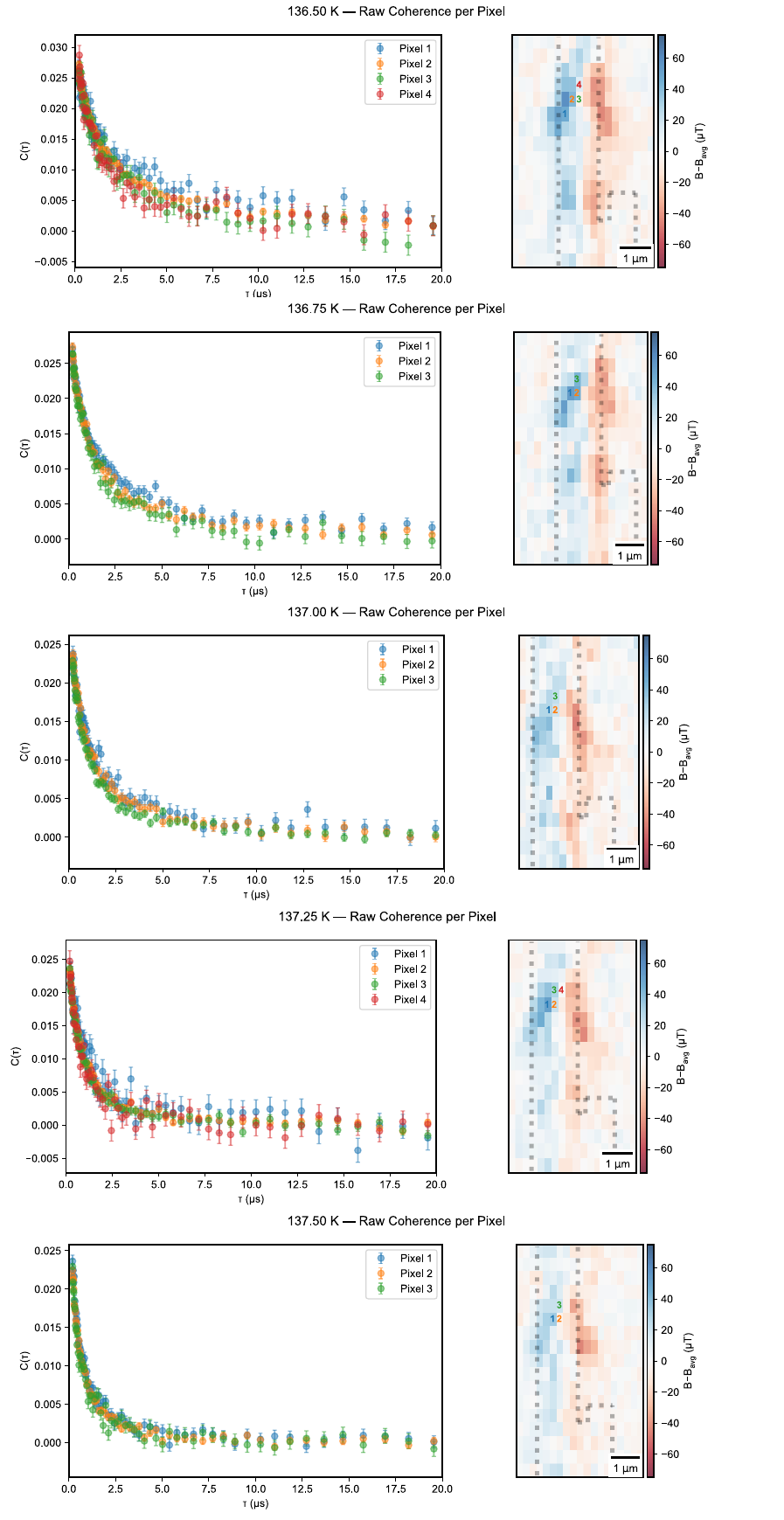}
\caption{\textbf{Pixel-to-pixel variation in the spin-echo coherence near criticality.} Raw coherence decays $C(\tau, T)$ resolved per pixel for several temperatures in the critical range (136.50--137.50~K). For each temperature, the left panel shows $C(\tau)$ for every tracked pixel (colored markers) with error bars, and the right panel shows the corresponding stray field image with pixel positions indicated by color-matched labels (each pixel $\sim$500~nm).}
\label{fig:SM_SpatialAnalysis}
\end{figure}

\begin{figure}
    \centering
    \includegraphics[width=\textwidth]{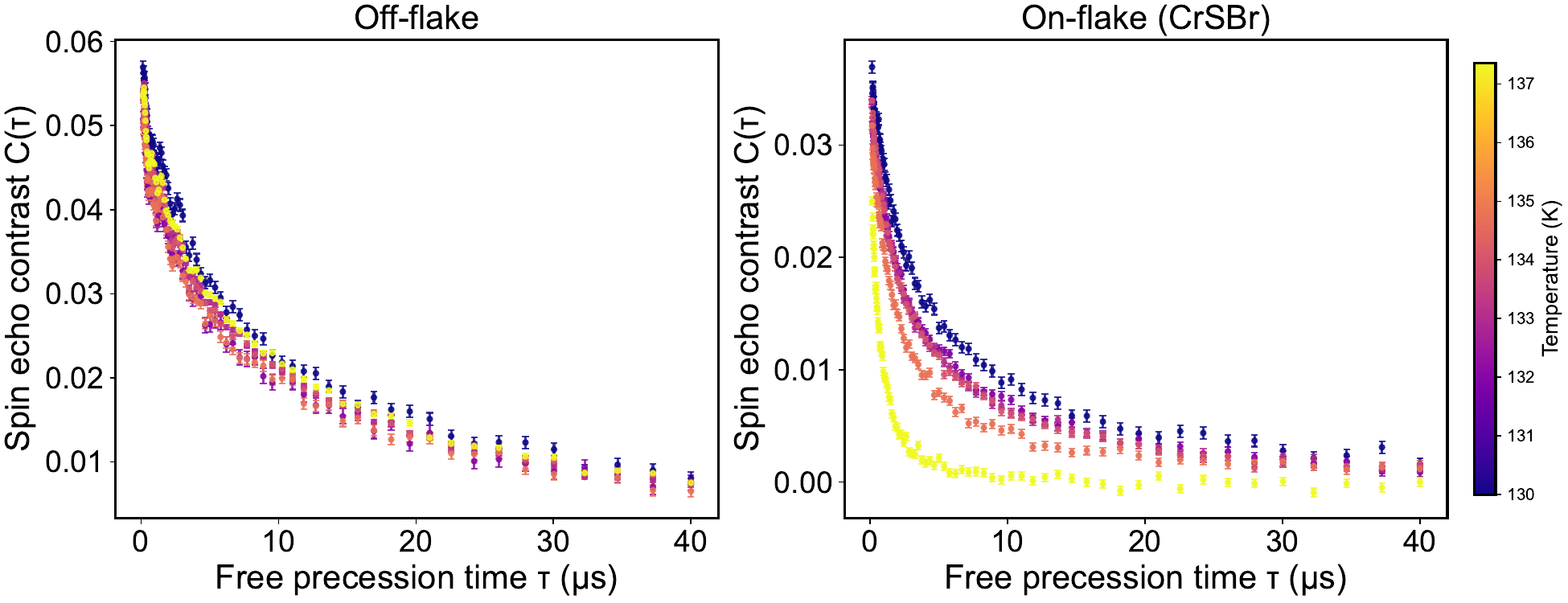}
    \caption{\textbf{Off-flake versus on-flake spin-echo coherence across the critical regime.} Spin-echo coherence decays $C(\tau)$ measured at a position off the CrSBr flake (left) and on the flake (right) at temperatures spanning the onset of critical fluctuations. The off-flake coherence is temperature-independent across the full range, reflecting the bare diamond spin-bath background decoupled from CrSBr. By contrast, the on-flake coherence shows a pronounced temperature dependence, with faster decoherence as the temperature approaches $T_C$. This confirms that the enhancement of NV decoherence in the critical regime originates from magnetic fluctuations in the CrSBr flake and not from any intrinsic temperature dependence of the diamond spin bath.}

    \label{fig:SM_OffOnFlakeComparison}
\end{figure}

\end{document}